\documentclass[aps,pre,twocolumn,superscriptaddress]{revtex4-2}

\usepackage{graphicx}
\usepackage{subfigure}
\usepackage{color}
\usepackage[dvipsnames]{xcolor}
\usepackage{amsmath,amssymb}
\usepackage[linktocpage,colorlinks=true,linkcolor=blue,citecolor=blue,breaklinks=true,urlcolor=blue]{hyperref}
\usepackage{soul}
\usepackage{cancel} 

\newcommand{\fig}[1]{Fig.~\ref{#1}}

\newcommand{\eq}[1]{Eq.~\ref{#1}}
\newcommand{\EQ}[1]{Equation~\ref{#1}}

\newcommand{\sctn}[1]{\S~\ref{#1}}
\newcommand{\movie}[1]{Movie~\ref{#1}}

\renewcommand{\vec}[1]{\boldsymbol{#1}}
\newcommand{\tens}[1]{\underline{\underline{#1}}}

\newcommand{\grad}{\vec{\nabla}}
\newcommand{\laplace}{\nabla^2}
\newcommand{\kb}{k_\textmd{B}}
\newcommand{\kbt}{\kb T}

\newcommand{\av}[1]{\left\langle #1 \right\rangle}

\newcommand{\abs}[1]{\left| #1 \right|}

\newcommand{\nn}{\nonumber}
\newcommand{\dt}{\delta t}
\newcommand{\Tint}{\mathfrak{T}}

\newcommand{\vel}{v}

\newcommand{\vcm}[1]{\vec{\vel}^{\textmd{cm}}_{#1}}
\newcommand{\velcm}[1]{\vel^{\textmd{cm}}_{#1}}
\newcommand{\hatcm}[1]{\hat{\vec{\vel}}^{\textmd{cm}}_{#1}}

\newcommand{\ori}{e}
\newcommand{\NCell}{N_{C}}  

\newcounter{movie}
\newcommand{\dummymov}[1]{\refstepcounter{movie}\label{#1}}

\newcommand{\ue}{School of Physics and Astronomy, The University of Edinburgh, Peter Guthrie Tait Road, Edinburgh, EH9 3FD, United Kingdom}
\newcommand{\puc}{Facultad de Física, Pontificia Universidad Católica de Chile, Santiago, 7820436, Chile}

\definecolor{pumpkin}{rgb}{1.0,0.4,0.0}
\definecolor{mygreen}{rgb}{0.0,0.55,0.3}
\definecolor{strawberry}{rgb}{1.0,0.0,0.5}
\definecolor{midnight}{rgb}{0.003921569,0.098039216,0.576470588}
\definecolor{saphire}{rgb}{0.0,0.196,0.372549}
\definecolor{crimson}{rgb}{0.75686,0,0.262745}
\definecolor{capri}{rgb}{0.0,0.768627,0.8745098}

\newcommand{\correctText}[2]{#2}
\newcommand{\correctTitle}[2]{#2}
\newcommand{\correctMath}[2]{#2}

\newcommand{\tocite}[1]{\textcolor{cyan}{\textsuperscript{*}}}

\begin{document}

\title{Multi-Particle Collision Framework for Active Polar Fluids}

\author{Oleksandr Baziei}
\affiliation{\ue}
\author{Benjamín Loewe}
\affiliation{\puc}
\author{Tyler N. Shendruk}
\affiliation{\ue}
\email{t.shendruk@ed.ac.uk}

\begin{abstract}
    \noindent
    Sufficiently dense intrinsically out-of-equilibrium suspensions, such as those observed in biological systems, can be modelled as active fluids characterised by their orientational symmetry. 
    While mesoscale numerical approaches to active nematic fluids have been developed, polar fluids are simulated as either ensembles of microscopic self-propelled particles or continuous hydrodynamic-scale equations of motion. 
    To better simulate active polar fluids in complex geometries or as a solvent for suspensions, mesoscale numerical approaches are needed. 
    In this work, the coarse-graining Multi-Particle Collision Dynamics (MPCD) framework is applied to three active particle models to produce mesoscale simulations of polar active fluids. 
    The first active-polar MPCD (AP-MPCD) is a variant of the Vicsek model, while the second and third variants allow the speed of the particles to relax towards a self-propulsion speed subject to Andersen and Langevin thermostats, respectively. 
    Each of these AP-MPCD variants exhibit a flocking transition at a critical activity and banding in the vicinity of the transition point. 
    We leverage the mesoscale nature of AP-MPCD to explore flocking in the presence of external fields, which destroys banding, and anisotropic obstacles, which act as a ratchet that biases the flocking direction. 
    These results demonstrate the capacity of AP-MPCD to capture the known phenomenology of polar active suspensions, and its versatility to study active polar fluids in complex scenarios.
\end{abstract}

\maketitle

\section{Introduction}
Active fluids spontaneously flow because they are composed of a continuum of microscopic motors that everywhere convert internal fuel into mechanical work~\cite{Alert1995, Ramaswamy2019} and serve the biophysics community as idealised generic models of collective motion. 
Biological examples of systems modelled as active fluids include suspensions of swimming bacteria~\cite{Aranson2022}, mitotic spindles~\cite{Conway2022}, epithelial tissues~\cite{Thuan2017}, ensembles of neural progenitor cells~\cite{Kawaguchi2017}, ectodermal cells~\cite{Yonit2021} and more~\cite{Lakshmi2022, Needleman2017}. 
Modelling such systems has primarily relied on two numerical approaches: particle-based approaches~\cite{Markus2020} and continuum theories~\cite{Marchetti2013}. 
These two limiting approaches are connected through the coarse-graining of self-propelled particle models into continuum theories, which is well established for standard models~\cite{Shaebani2020}. 
Indeed, Vicsek's original model of self-propelled particles with minimal interactions leading to collective flocking~\cite{vicsekspp} was immediately followed by the Toner-Tu model~\cite{TonerTu1995}, which applied symmetry arguments rather than direct coarse-graining to arrive at a continuum theory, the renormalisation of which can predict the critical behaviour~\cite{Chate2024}.
Both the particle-based Vicsek model (and its variants~\cite{Chate2020, Patelli2019}) and the Toner-Tu model~\cite{Reinken2022, Haoran2024,  Kiran2023, James2021} (often supplemented by a Swift-Honenberg-type ansatz~\cite{Wensink2012, Oza2016}) have been remarkably fruitful models of active matter. 

However, particle-based and continuum models represent the limiting approaches to simulating soft condensed matter systems~\cite{Slater2009}. 
They are supplemented by a third, middling approach: mesoscale simulations. 
Mesoscale lies midway between the microscopic scale of a material and the macroscopic scale of the system. 
Mesoscale simulations, such as coarse-grained molecular dynamics\cite{soumil2021,shi2023}, dissipative particle dynamics~\cite{Espanol2017,santo2021,wang2021} and multi-particle collision dynamics~\cite{howard2019, Lakshmi2022}, have proven valuable for simulating soft condensed matter systems. 
In particular, systems involving mesoscale solutes, such as anisotropic colloids~\cite{Zablotsky2017,Yamanouchi2021,Satoh2022,DuarteAlaniz2023}, polymers~\cite{Sappl2023}, or complex fluids~\cite{dicintio2015,mandal2019,toneian2019,reyesArango2020,ilg2022,ilg2022b} are amenable to mesoscale methods. 
Likewise, particle-based coarse-grained approaches can more easily simulate complex geometries, such as wavy channels~\cite{wamsler2024}, funnels~\cite{wen2022} or confined swimmers~\cite{zottl2018}, because they can avoid complex meshing methods that are often crucial to solving continuum differential equations in the vicinity of boundaries~\cite{Kasun2023}.
However, few mesoscale techniques have been developed for active fluids, despite the importance of understanding the dynamics of passive inclusions embedded in active surroundings~\cite{ray2023,tayar2023} and active dynamics in complex geometries~\cite{Olsen2020, kuhn2021, McClure2022, martinez2018, Martinez2020, serna2022, Chen2022, vahabli2023, keogh2024} for designing material applications for active matter. 
To better simulate active polar fluids as a background active medium for dynamic mesoscale objects in complex geometries, coarse-grained numerical approaches are needed. 


This work introduces three mesoscale approaches to simulating active polar fluids. These mesoscale algorithms utilise the Multi-Particle Collision Dynamics (MPCD) framework, which we introduce in \sctn{sctn:MPCD}. 
We then review the Vicsek model (\sctn{sctn:vicsek-model}), before using it as the basis for the first active polar MPCD (AP-MPCD) in  \sctn{sctn:vicsekMPCD}, which replaces interactions with multi-particle collisions that preserve the speed of each fluid particle. 
This ``Vicsek-MPCD'' model exhibits a flocking transition at sufficiently high activity and density (\sctn{sctn:vicsek-MPCD}). 
\correctText{We show that the activity at which the system transitions to flocking follows the predicted low-densities scaling[57] and measure the high-density scaling behaviour.}{We demonstrate that the system transitions to flocking at an activity consistent with the predicted low-density scaling~\cite{Chate2008}. 
Furthermore, we go beyond the low-density limit and characterise the scaling behaviour at high densities.}

We then relax the constraint that fluid particles have a fixed self-propulsion speed through two thermostatted approaches. 
The first of these is an Andersen-thermostatted AP-MPCD (\sctn{sctn:andersen-vicsek}), while the second is Langevin\correctText{ }{-}thermostatted AP-MPCD (\sctn{sctn:langevin-vicsek}). 

After demonstrating that all three algorithms exhibit a flocking transition, we consider how the MPCD framework simulates collective motion in complex geometries focusing on one of the three algorithms. 
Concentrating on \correctText{L}{Langevin} AP-MPCD, we demonstrate the coexistence of a disorderly dilute phase and a dense flocking phase in system near the flocking transition, which is expressed via stable bands (\sctn{sctn:banding}). In addition, the influence of gravity on a system of self-propelled particles demonstrates that external forces suppress the phase transition (\sctn{sctn:gravity}). 
Finally in \sctn{sctn:obstacles}, we embed the polar fluid in geometries constructed of asymmetric obstacles. 
The active fluid inherits the broken symmetry of the obstacles, which not only sets the direction of flocking but impacts the flocking transition.

Through these results, we demonstrate that the AP-MPCD is a versatile numerical tool for simulating active polar fluids. 
The framework offers numerous tunable parameters, including the collision operator for particles and versatile boundary conditions. 
This flexibility allows for modelling of a variety of different scenarios, while reproducing known phenomena, such as order-disorder phase transition in the Vicsek model or banding in the vicinity of critical activity. In the future, its flexibility will allow it to be extended beyond self-propelled particles to simulate more complex biologically relevant active systems such as active suspensions \cite{SandS}.

\section{Methods}
To introduce novel active polar algorithms we have to discuss the pre-existing passive hydrodynamic solver, Multi-Particle Collision Dynamics (MPCD; \sctn{sctn:MPCD}). 
This computational technique can employ intrinsic thermostats (\sctn{sctn:AndersenMPCD} and \sctn{sctn:Langevin-MPCD}). 
After summarising MPCD, we review the Vicsek model (\sctn{sctn:vicsek-model}), before proposing hybrid and novel mesoscale algorithms for active polar fluids. 

\subsection{Multi-particle Collision Dynamics}
\label{sctn:MPCD}

Multi-Particle Collision Dynamics (MPCD) is a particle-based mesoscale simulation technique that intrinsically incorporates thermal fluctuations to model hydrodynamic flows~\cite{howard2019,Gompper2009,kapral2008,yeomans2006}. 
The MPCD algorithm is a two-step method composed of a \textit{streaming step} in which particles translate and a \textit{collision step} in which particles stochastically exchange properties with their neighbours while respecting appropriate conservation laws. In MPCD, the fluid is discretised into $N$ point particles of mass $m$. During the streaming step, the position of the $i$\textsuperscript{th} particle is updated as
\begin{equation}
\label{eq:stream}
  \vec{r}_i(t+\dt) = \vec{r}_i\left(t\right) + \vec{\vel}_i\left(t\right)\dt,
\end{equation}
in which $\vec{r}_i\left(t\right)$ is the instantaneous position, $\vec{\vel}_i\left(t\right)$ the instantaneous velocity and $\delta$t is the time step. 

\begin{figure}[tb]
    \centering
    \begin{subfigure}
        \centering
        \includegraphics[width=0.21\textwidth]{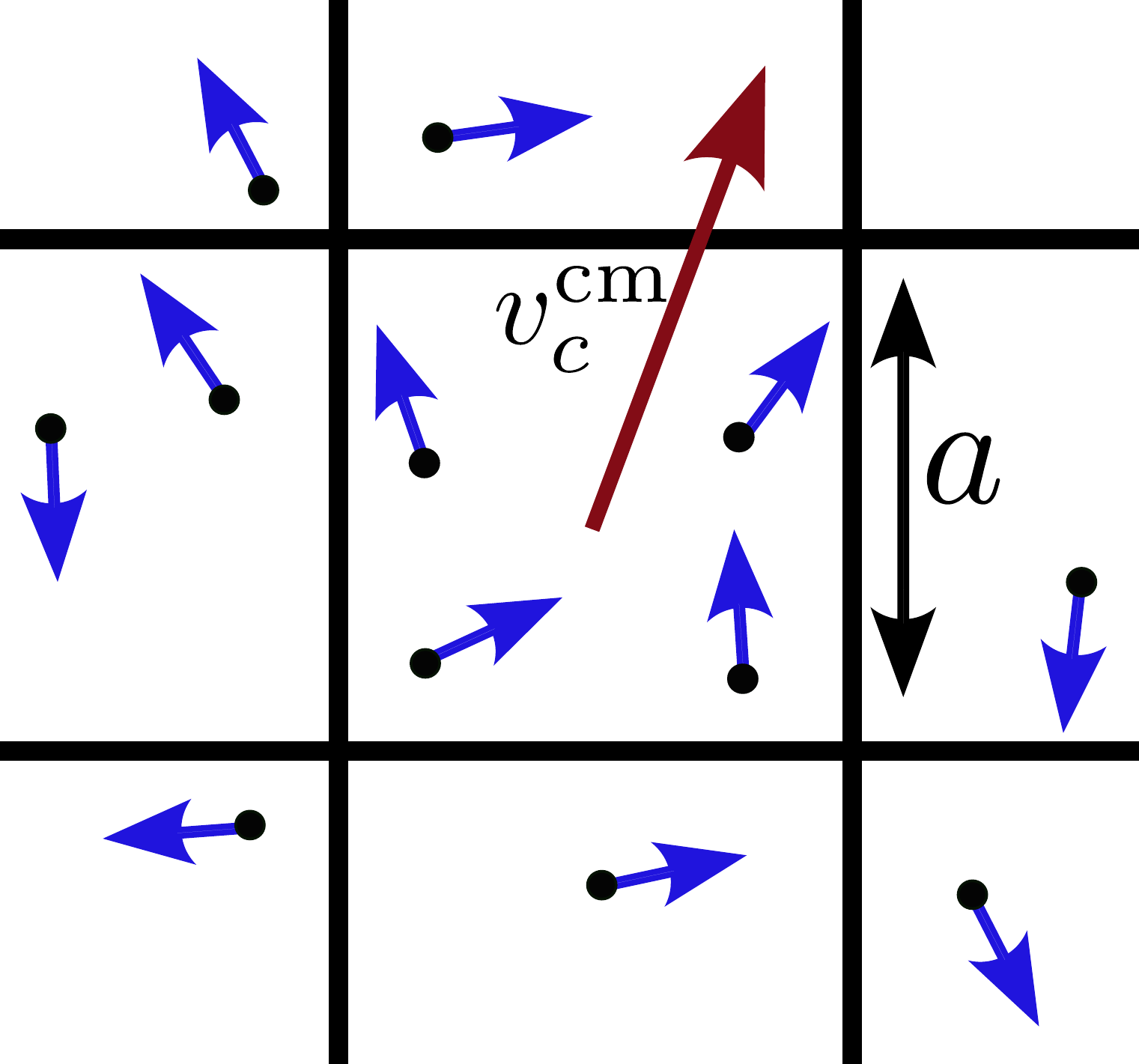}{(a)}
    \end{subfigure}
    \begin{subfigure}    
        \centering
        \includegraphics[width=0.2\textwidth]{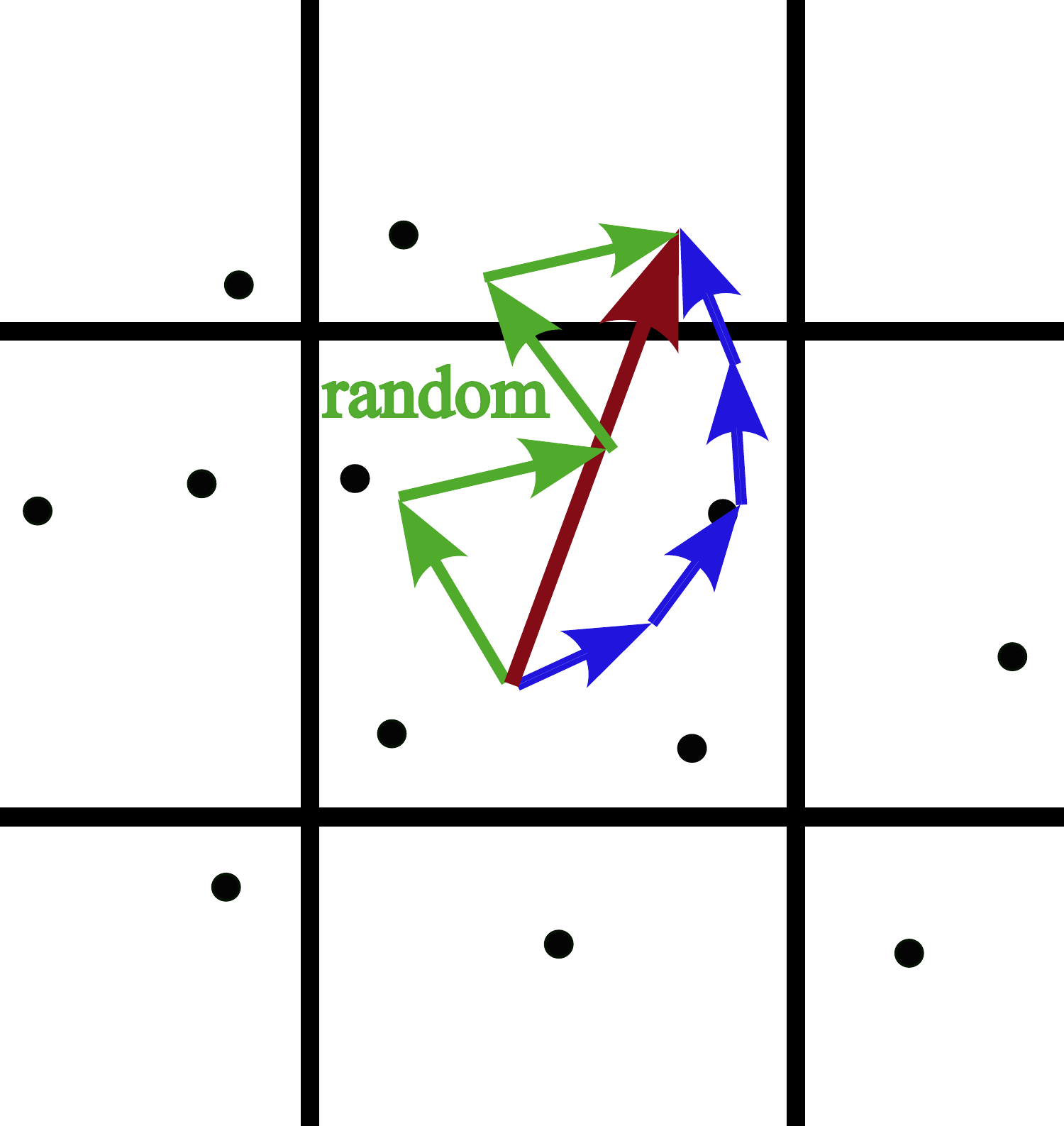}{(b)}
    \end{subfigure}
    \caption{
        \label{fig:MPCD} 
        Schematic of the MPCD framework. 
        (a) Space is divided into cells with linear size $a$. Velocities $\vec{\vel}_i\left(t\right)$ (blue arrows) of each particle sum up to cell $c$'s centre of mass velocity $\vcm{c}\left(t\right)$ (red arrow). 
        (b) A stochastic Andersen collision operation (\eq{eq:andersen}) that conserves momentum assigns random velocities (green arrows) to each particle in cell $c$. 
    }
\end{figure}

The system is of size $L$ and is divided into cells of size $a$, within which the collision steps occur (\fig{fig:MPCD}a). 
Each cell is indexed $c$. 
This division into cells significantly reduces the computational load by restricting interactions to particles within the same cell. A random grid shift ensures Galilean invariance~\cite{Ihle2001}. Only particles in the same cell interact and the interactions involve all particles in the same cell.
Within each cell $c$, the collision operator $\vec{R}_{ic}$ stochastically produces new velocities 
\begin{equation}
    \label{eq:vicsekcollision}
     \Vec{v}_i (t+\dt)= \vec{R}_{ic}\left(\vcm{c},\Vec{v}_i;t\right),
\end{equation}
for each of the $i$ particles in terms of their original velocity $\Vec{v}_i\left(t\right)$ and the centre of mass velocity of the local cell $\vcm{c}\left(t\right) = \av{ \Vec{v}_j }_c = \NCell^{-1} \Sigma_{i=1}^{\NCell} \vec{\vel}_i$, in which $\av{ \cdot }_c$ denotes the average over cell $c$ assuming all particles have the same mass $m$ and $\NCell$ is the number of particle in cell $c$. 
As the simulation progresses, particle velocities are updated based on the collision operator, introducing randomness and mimicking particle interactions. Although there are several choices for collision operator~\cite{Strobl2017}, this study focuses on two thermostatted approaches: Andersen MPCD and Langevin MPCD~\cite{Noguchi2007}.

\subsubsection{Andersen MPCD}\label{sctn:AndersenMPCD}
Thermostatted MPCD collision operators maintain constant temperature conditions~\cite{Noguchi2007}.
Though momentum is conserved, the energy is not; it is constantly fluctuating about a constant thermal energy $\kbt$. 
Thermostatted collision operators include thermostatting inside each cell, such that no global velocity rescaling is required, even if external fields are applied. 
Andersen-thermostatted MPCD implements an Andersen thermostat within the collision operation while locally conserving momentum. 
During the collision steps (\fig{fig:MPCD}b), each particle is reassigned its velocity with a random value drawn from a Maxwell–Boltzmann distribution with energy $\kbt$ in an isotropically random direction while preserving momentum by subtracting the average of all random velocities. 
The collision operator is 
\begin{equation}
  \label{eq:andersen}
   \vec{R}_{ic}\left(\vcm{c},\Vec{v}_i;t\right) = \vcm{c}\left(t\right) + \vec{\xi}_i - \av{ \vec{\xi} }_c ,
\end{equation}
in which $\vec{\xi}_i$ are random velocities components drawn from a Gaussian distribution \correctText{centred on zero, with thermodynamic mean $\av{ \vec{\xi}_i\left(t\right) } = 0$}{with zero thermodynamic mean} and \correctText{variance $\av{ \vec{\xi}_i\left(t\right)\vec{\xi}_j(t^\prime) } = \delta_{ij}\delta(t-t^\prime)\tens{I}$ for Kronecker delta $\delta_{ij}$, Dirac delta $\delta(t-t^\prime)$ and identity matrix $\tens{I}$}{uncorrelated variance}.
The instantaneous cell-average of random velocities is $\av{ \vec{\xi} }_c$. 
Andersen MPCD has found widespread success in simulating situations in which energy is injected into the system from an external force \cite{Noguchi2008, Noguchi2007, mandal2019}.

In this work, particle mass $m = 1$, energy $\kbt = 1$ and cell size $a = 1$ set the simulation units. 
The unit time is $\correctMath{}{\Delta t_\text{MPCD}=} a\sqrt{m/\kbt} = 1$ and the time step is $\dt = 0.1\correctMath{}{\Delta t_\text{MPCD}}$. 
The average number of particles per cell $\av{N_c}$ is equivalent to the particle mass density in the simulations. 
All results are reported in simulation units and all the simulations are in $D=2$ dimensions. 
\correctText{}{The random variables are drawn from Gaussian distribution with $\av{ \vec{\xi}_i\left(t\right) } = 0$ and $\av{ \vec{\xi}_i\left(t\right)\vec{\xi}_j(t^\prime) } = \delta_{ij}\delta(t-t^\prime)\tens{I}$ for Kronecker delta $\delta_{ij}$, Dirac delta $\delta(t-t^\prime)$ and identity matrix $\tens{I}$.}
The total number of particles is $N=\av{N_c} L^2$. 
Unless stated otherwise, the system is square and has a size $L = 100$ with periodic boundary conditions in all Cartesian directions.
A warmup time of $\correctMath{T}{\Tint}_W = 1.5 \times 10^4 \dt$ is found to be sufficiently long for all systems to reach steady state. 
After the warmup, data collection occurs for $\correctMath{T}{\Tint}=5 \times 10^5 \dt$. 

\subsubsection{Langevin MPCD}\label{sctn:Langevin-MPCD}

The Langevin collision operator is another approach to thermostatting MPCD~\cite{Noguchi2007}. In Langevin MPCD, the collision operator is
\begin{align}
    \vec{R}_{ic}\left(\vcm{c},\Vec{v}_i;t\right) &= \vcm{c}\left(t\right) + b\left[\vec{\vel}_i\left(t\right) - \vcm{c}\left(t\right)\right] \nn\\
    &\qquad + \ell\left( \vec{\xi}_i - \Sigma_{j}^{\NCell} \frac{\vec{\xi}_i}{\NCell} \right),
    \label{eq:LangevinPassive}
\end{align}
in which
\begin{align}
  b &= \frac{2m- \gamma \dt}{2m + \gamma \dt} \label{eq:b}, \\
  \ell &= \frac{2\sqrt{\gamma \dt}}{2m + \gamma \dt} \label{eq:ell}
\end{align}
with $\gamma$ a drag coefficient. The collision operator exerts two effects on the fluid particles: a viscous force that opposes the relative motion of fluid particles within each cell, and random kicks resulting from the thermal bath, described by the white noise $\vec{\xi}_i$ with all the same properties as in \sctn{sctn:AndersenMPCD}. 
The drag factor (\eq{eq:b}) and the noise strength (\eq{eq:ell}) are related through the fluctuation-dissipation theorem.

The units are the same as in Andersen\correctText{-}{ }MPCD (\sctn{sctn:AndersenMPCD}) but include the additional parameter $\gamma$. 
Setting $\gamma = 2m /\dt$ and $m=1/2$ causes $b=0$ and $l=1$, which reduces Langevin MPCD (\eq{eq:LangevinPassive}) to Andersen MPCD (\eq{eq:andersen}). 
Here, however, the drag is set to $\gamma=1$, which is much less than $2m/\dt = 20$ and gives $b=0.905$ and $\ell=0.301$. 
Although Langevin\correctText{-}{ }MPCD is a simple model for including dissipation, it is employed less frequently in the literature than Andersen\correctText{-}{ }MPCD. 

In this work, we propose an additional active polar term in $\vec{R}_{ic}$ for both Andersen\correctText{-}{} and Langevin\correctText{-}{ }MPCD algorithms. 
The active term is proposed after implementing the Vicsek model in the MPCD framework to produce a hybrid Vicsek-MPCD algorithm. 
Before proposing a hybrid Vicsek-MPCD approach, we review the traditional Vicsek model. 

\begin{figure}[tb]
    \centering
    \includegraphics[width=0.4\textwidth]{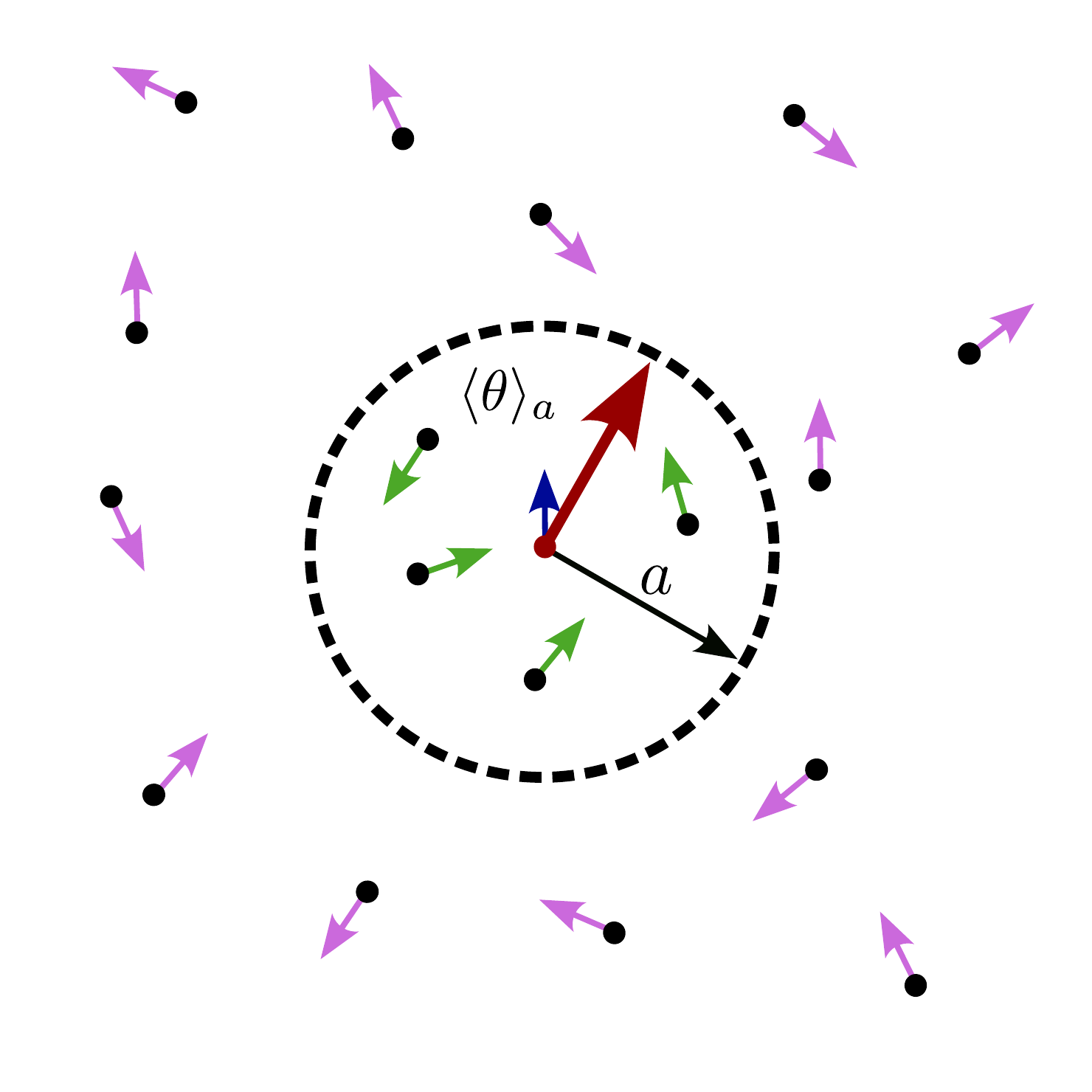}
    \caption{
        \label{fig:VicsekModel} 
        Schematic of the Vicsek model. 
        Each particle $i$ has an instantaneous velocity (blue arrow).
        The average direction $\av{\theta}_a$ within the interaction distance $a$ around particle $i$ is given by the red arrow.
        Green arrows represent instantaneous velocities of the particles within the interaction range, while magenta arrows represent those outside the interaction range.
        }
\end{figure}

\subsection{Vicsek Model}
\label{sctn:vicsek-model}
The Vicsek model operates on a simple rule: \textit{At each time step, each self-propelled particle attempts to align its direction of motion with the average direction of its neighbours while subjected to stochastic perturbations.}
This models polar interactions between self-propelled particles~\cite{vicsekspp}.
It is a ``dry'' active model~\cite{Marchetti2013} that conserves energy since the particles travel at a constant speed. 
The alignment interaction, however, is active, injecting momentum through the reorientation of velocities. 
It offers a simplified-yet-insightful framework that captures rich collective dynamics in a number of different contexts~\cite{Fazli2021, Kreienkamp2022}. 
In particular, it exhibits a collective transition to a cohesive flocking state, in which individual particles collectively converge upon a spontaneously chosen direction~\cite{ginelli2016}. 
This manifests as a kinetic transition through spontaneous rotational symmetry breaking~\cite{Puzzo2022, Wu2021, Chatterjee2022}. 

In the original Vicsek model, the position of the $i$\textsuperscript{th} particle is updated as
\begin{equation}
\label{eq:VicsekStream}
  \vec{r}_i(t+\dt)=\vec{r}_i\left(t\right)+\vec{\vel}_i\left(t\right)\dt, 
\end{equation}
in which $\vec{\vel}_i \left(t\right)=\vel_0 \vec{\ori}_i\left(t\right)$ is the velocity of the $i$\textsuperscript{th} particle, which maintains a constant speed $\vel_0$, but has an instantaneous orientation $\vec{\ori}_i\left(t\right) = \left( \cos\theta_i\left(t\right),\sin\theta_i\left(t\right) \right)$ determined by the angle $\theta_i \left(t\right)$ in 2D --- though Vicsek model can also be implemented in 3D~\cite{Wu2021}.

\begin{figure}[tb]
    \centering
    \includegraphics[width=0.25\textwidth]{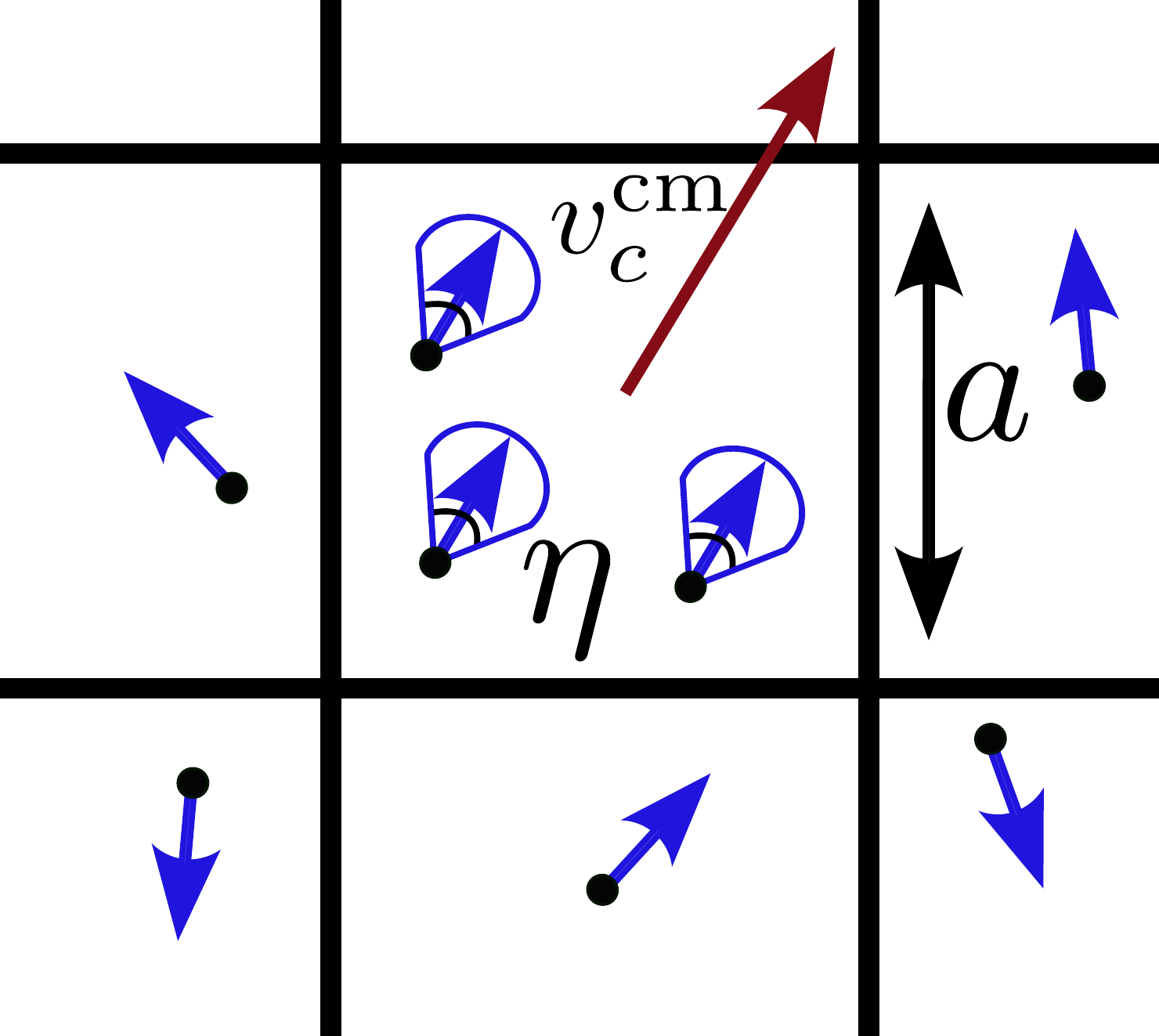}
    \caption{
        \label{fig:VicsekModelSchematic} 
        Schematic of the hybrid  Vicsek-MPCD algorithm. 
        The collision operator in the Vicsek-MPCD model (\eq{eq:vicsekHybrid}) rotates velocity of the particle while speed remains unchanged. 
        The red arrow is the centre of mass velocity $\vcm{c}$ in a cell of size $a$. 
        Blue arrows are the instantaneous velocity of each particle. 
        Cones represent the distribution of random velocities, characterized by the strength of the noise $\eta$.
        }
\end{figure}

Between time steps, each particle instantaneously updates its heading $\vec{e}_i$ based on the instantaneous state of its neighbours (\fig{fig:VicsekModel}).
The direction of motion of all the particles are updated simultaneously at each time step according to
\begin{equation}
\label{eq:angle}
  \theta_i (t+\dt)=\av{ \theta(t) }_{r_{ij} \leq a} + \zeta_i.
\end{equation}
The first term, $\av{ \theta (t)}_{r_{ij} \leq a} = \arctan[ \av{\sin\theta(t)}_{r_{ij} \leq a} / \av{\cos\theta(t)}_{r_{ij} \leq a}]$, represents the local direction averaged over all particles within the interaction range of radius $a$ (\fig{fig:VicsekModel}).
The average is centred on particle $i$ and includes particle $i$, as well as all particles $j$ for which $r_{ij} = \sqrt{\left(\vec{r}_i-\vec{r}_j\right)^2} \leq a$. 
The second term is a noise $\zeta_i$, which is a random angle drawn uniformly from the interval $[- \pi\eta, \pi\eta]$. The factor $\eta$ represents the width of the angular distribution and thus sets the strength of the noise. We will often use the activity parameter
\begin{equation}
 \label{eq:activity}
  \alpha = 1 - \eta.
\end{equation}
There are six parameters in the Vicsek model: $a$, $\dt$, $\vel_0$, $\eta$, $L$, and $N$. 
The first two set the units of length ($a$) and time ($\dt$). 
Consequently, the three adjustable parameters are $\vel_0$, $\eta$ and the particle density $\rho=N/L^2$.

To quantify the degree of flocking, the normalised ensemble averaged speed
 \begin{equation}
  \phi = \frac{\abs{ \av{\vec{v}} }}{v_0}=\frac{1}{N \vel_0} \abs{ \Sigma_{i=1}^{N} \vec{\vel}_i }
  \label{eq:orderParam}
\end{equation}
is used as the order parameter. 
When there is no order in the system, the particles move in independent directions and there is no net transport, such that this flocking order parameter goes to zero ($\phi\to 0$). 
However, as the system self-organises into a flock, particles collectively move in a spontaneously chosen direction and $\phi\to1$. 

The continuum description of the Vicsek model is given by the Toner-Tu equation~\cite{TonerTu1995}
 \begin{equation}
  (\partial_t+\lambda\vec{v}\cdot\grad)\vec{v}+\grad p = \alpha^\prime\vec{v}+\nu\laplace\vec{v}-\beta \vec{v} \abs{\vec{v}}^2,
  \label{eq:tonerTu}
\end{equation}
in which $\alpha^\prime$ and $\beta$ are fixed parameters and $\vec{v}(\vec{r},t)$ satisfies the incompressibility condition $\vec{\nabla}\cdot \vec{v} = 0$. 
\EQ{eq:tonerTu} is a continuum hydrodynamic model that describes the dynamics of self-propelled particle systems exhibiting ordered collective motion. 
In conjunction with the Swift-Hohenberg ansatz~\cite{Wensink2012, Oza2016}, \eq{eq:tonerTu} is commonly used as continuous model of bacterial turbulence \cite{Xia}.

While the Vicsek model is microscopic and the Toner-Tu equation is a macroscopic hydrodynamic-scale model, mesoscale simulations are lacking. 
We now propose a hybrid MPCD technique for mesoscale simulations of active polar fluids to fill this gap.

\section{Vicsek-MPCD Hybrid Algorithm}
\label{sctn:vicsekMPCD}

Both the Vicsek model and the MPCD algorithm are composed of two steps: streaming and collision. The streaming step is identical in the two models (\eq{eq:stream} and \eq{eq:VicsekStream}). 
In the Vicsek model, the alignment of particle $i$ with the particles within the interaction radius acts as a ``collision'' event, while an MPCD collision is a stochastic local exchange within each lattice-based cell. 
In the following section, we implement Vicsek alignment within MPCD cells to create a hybrid Vicsek-MPCD algorithm. 

\begin{figure}[tb]
    \centering
    \includegraphics[width=0.5\textwidth]{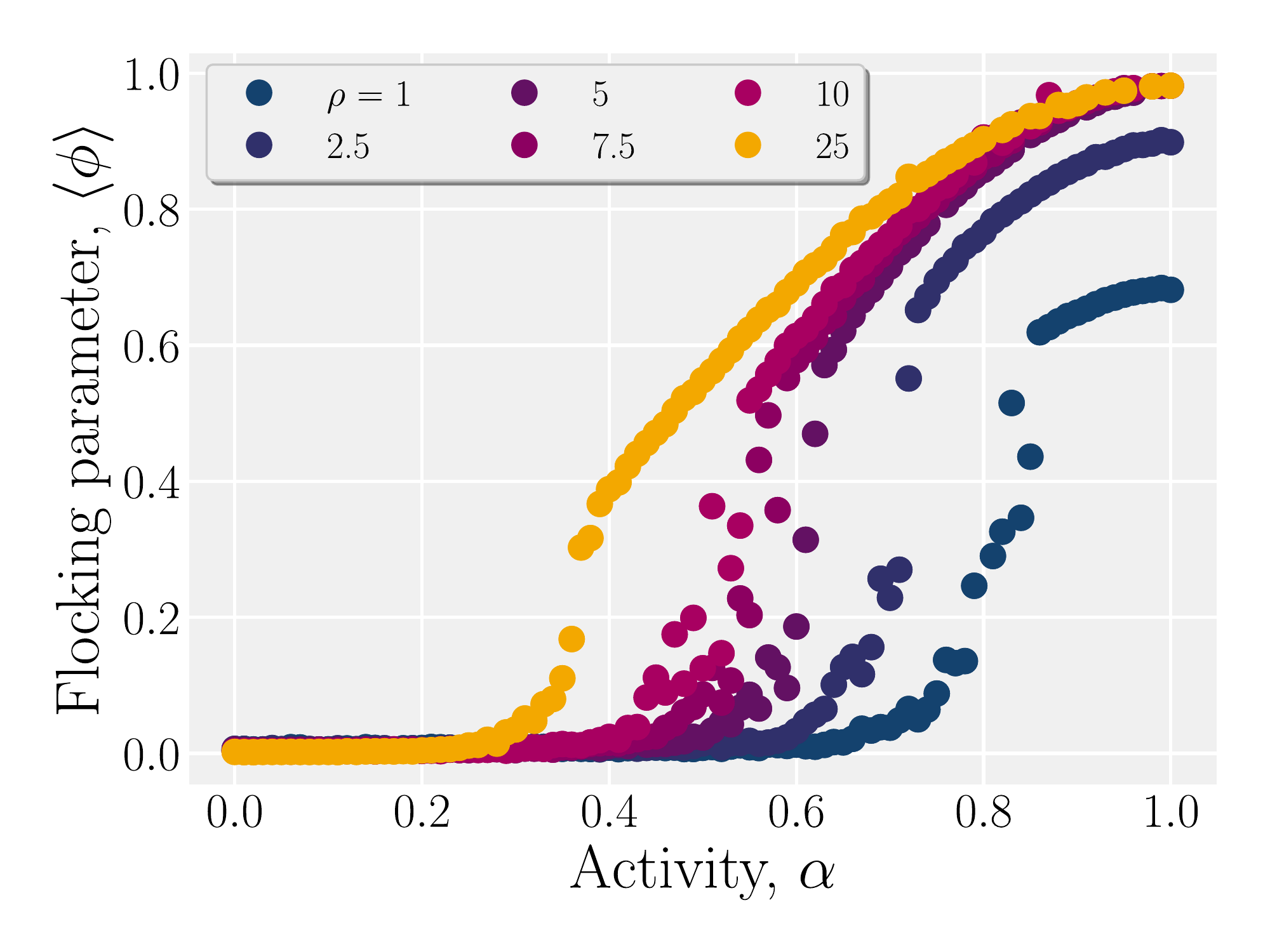}
    \caption{
        \label{fig:VicsekModelFigure}
        Flocking transition in Vicsek-MPCD for various densities $\rho$ as a function of activity $\alpha = 1 - \eta$. 
        The time-averaged flocking order parameter $\av{\phi}$ is the magnitude of the average velocity normalised by $v_0=\sqrt{2}$ (\eq{eq:orderParam}).
        Each point represents the average \correctText{}{of} an entire simulation.
    }
\end{figure}

\subsection{Vicsek-MPCD Collision Operator}
\label{sctn:vicsek-MPCD}
The proposed Vicsek model builds on the MPCD framework while preserving the original Vicsek concept of aligning particles with constant speed. This amounts to an operation that rotates the original velocity $\Vec{v}_i\left(t\right)$ to align with the local average $\vcm{c}$ plus some noise (\fig{fig:VicsekModelSchematic}). 
The collision operator that updates the velocities $\Vec{v}_i (t+\dt)= \vec{R}_{ic}\left(\vcm{c},\Vec{v}_i;t\right)$ (\eq{eq:vicsekcollision}) for the Vicsek-MPCD algorithm is
\begin{align}
    \label{eq:vicsekHybrid}
    \vec{R}_{ic}\left(\vcm{c},\Vec{v}_i;t\right) &= \mathcal{R}\cdot\Vec{v}_i\left(t\right), \\
    \mathcal{R} &= 
    \begin{bmatrix} 
        \cos\left(\delta \theta_i\right) & -\sin\left(\delta \theta_i\right) \\ 
        \sin\left(\delta \theta_i\right) & \cos\left(\delta \theta_i\right) 
    \end{bmatrix},
\end{align}
in which each particle\textquotesingle s velocity is rotated at each time step over the angle
\begin{align}
    \delta \theta_i\left(t\right) &= \av{\theta}_c  - \theta_i\left(t\right) + \zeta_i.
\end{align}
This is identical to the Vicsek model (\eq{eq:angle}), except that the average orientation is taken over the MPCD cells, rather than over the interaction range $a$. 
The noise $\zeta_i$ is drawn from the same uniform distribution of width  $[- \pi\eta, \pi\eta]$ and activity is $\alpha = 1 - \eta$. 
The Vicsek order parameter $\phi=\abs{\av{\vec{v}}} / v_0$ (\eq{eq:orderParam}) applies to the Vicsek-MPCD algorithm.
The 
particle velocity components are initialised randomly from a uniform distribution 
\begin{equation}
    p\left(v_{i,j}\right) = \sqrt{ k_BT/m }(2 v_{i,j} \varrho - 1), 
    \label{eq:initDist}
\end{equation}
in which $j\in\{x,y\}$ is the spatial component and $\varrho$ is randomly drawn uniformly from the interval $[0, 1)$. This distribution is characterised by the average speed
\begin{equation}
    \label{eq:vicsekDist}
    \vel_0 = \sqrt{\frac{2k_BT}{m}}.
\end{equation}
Since $k_BT = 1$ and $m = 1$, the characteristic speed is $\vel_0 = \sqrt{2}$.
The flocking order parameter fluctuates in time and so is averaged \correctText{$\av{\phi}=T^{-1} \sum_{t=0}^T \phi\left(t\right)$}{$\av{\phi}=\correctMath{T}{\Tint}^{-1} \sum_{t=0}^{\correctMath{T}{\Tint}}\phi\left(t\right)$}. 
The main difference between the traditional Vicsek model and the proposed Vicsek-MPCD is that the interactions do not happen within an interaction range around each individual particle but rather happen in multi-particle collision events.
We will show that this algorithm reproduces the main properties of the traditional Vicsek model.

\subsection{Vicsek-MPCD Results}

We simulate a series of different particle densities 
$\rho = N / L^2 \in \left[ 0.001,25\right]$ and activities in the interval $\alpha \in [0, 1]$. 
The initial condition has $N$ particles uniformly distributed within the system with their velocities drawn from \eq{eq:initDist}. 

\begin{figure}[tb]
    \centering
    \includegraphics[width=0.5\textwidth]{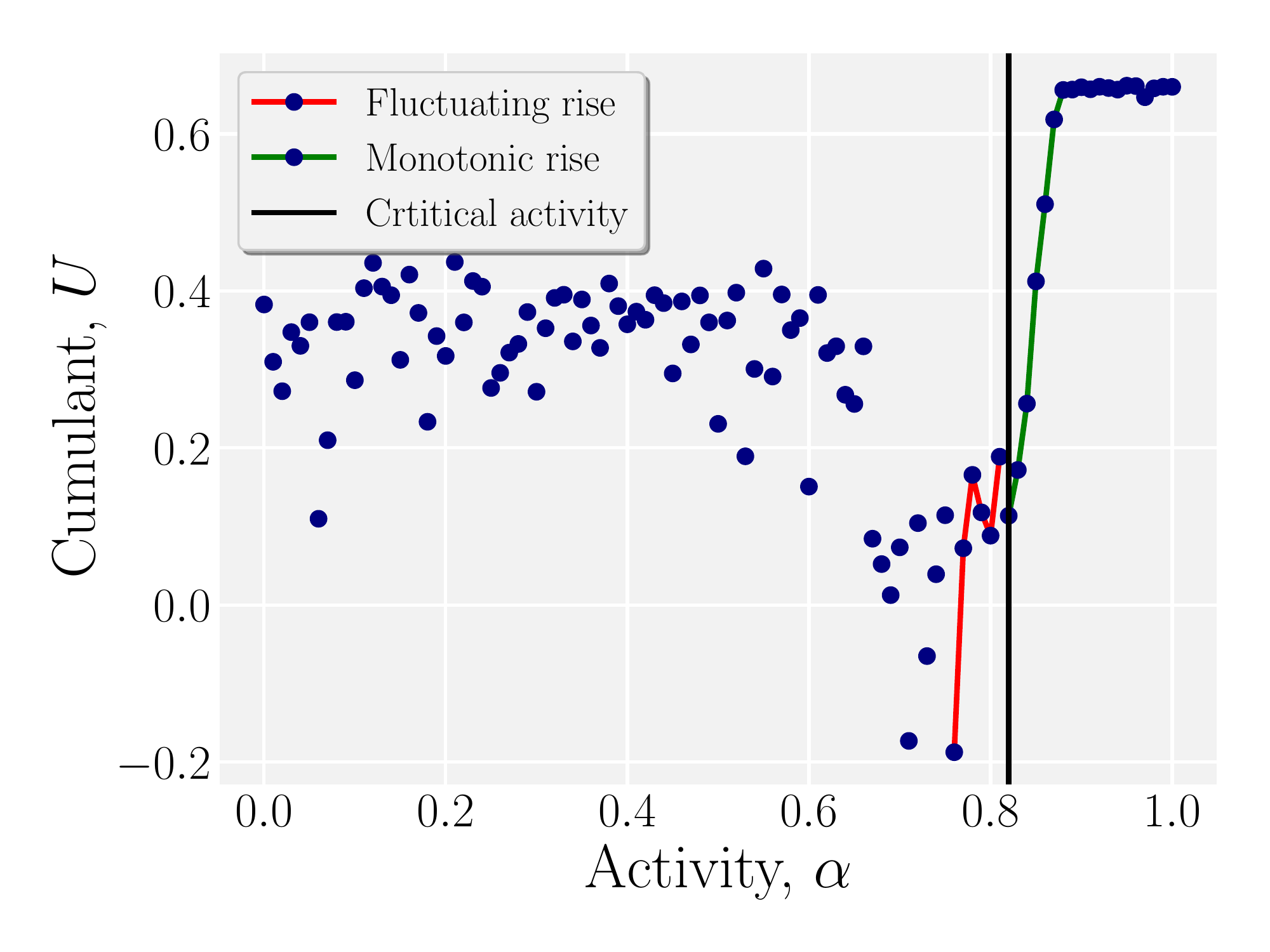}
    \caption{
        \label{fig:BinderCumulant} 
        Binder cumulant $U$ (\eq{eq:Binder}) for density $\rho = 1$ at different activities $\alpha$. 
        The rapid change in $U$ between $0.6 < \alpha < 0.9$ indicates a transition. 
        The critical activity $\alpha_k = 0.82$ (vertical black line) is identified as the point between fluctuations immediately after the global minimum (red line) and the monotonic rise (green line). 
    }
\end{figure}

The hybrid Vicsek-MPCD algorithm exhibits a flocking transition (\fig{fig:VicsekModelFigure}). 
For sufficiently small activities $\alpha = 1 - \eta$, the temporally averaged flocking order parameter $\av{\phi}$ is zero \correctText{}{(\movie{mov:disordered})}; however, after some critical activity $\alpha_k$ which depends on the particle density, the average speed starts to increase. 
At higher activities, the order parameter plateaus to $\av{\phi} \to 1$ \correctText{}{(\movie{mov:ordered})}. 
The denser the system, the lower the activity at which the ensemble becomes ordered. 
In a minority of numerical realisations, the system may remain in the disordered state for the duration of the simulation, without finding the expected steady-state flocking (\textit{e.g.} \fig{fig:VicsekModelFigure}; $\rho=5$ and $\alpha=0.75$). 

To measure the critical activity $\alpha_k$, we employ a Binder cumulant~\cite{Binder1981}
\begin{equation}
    \label{eq:Binder}
    U = 1 - \frac{ \av{\phi^4} }{ 3\av{\phi^2}^2},
\end{equation}
which is the fourth-order cumulant of the order parameter. 
The Binder cumulant quantifies how much a distribution differs from a normal one~\cite{Binder2010}. 
In the vicinity of the critical point, order parameters generally exhibit broader than normal distributions due to their scale invariance at the critical point ~\cite{Binder1997}, causing rapid changes in $U$ in the vicinity of the critical activity (\fig{fig:BinderCumulant}). 
This property can be employed to identify the critical activity $\alpha_k$ at which ordered flocking emerges more precisely than directly from \fig{fig:VicsekModelFigure}. 
The negative dip suggests coexistence and a first order phase transition \cite{Iino2019}. 
The critical activity $\alpha_k$ is identified as the value of $\alpha$ after the global minimum no-longer stochastically fluctuates but rather steadily rises monotonically, until $U$ plateaus.

\begin{figure}[tb]
    \centering
    \includegraphics[width=0.5\textwidth]{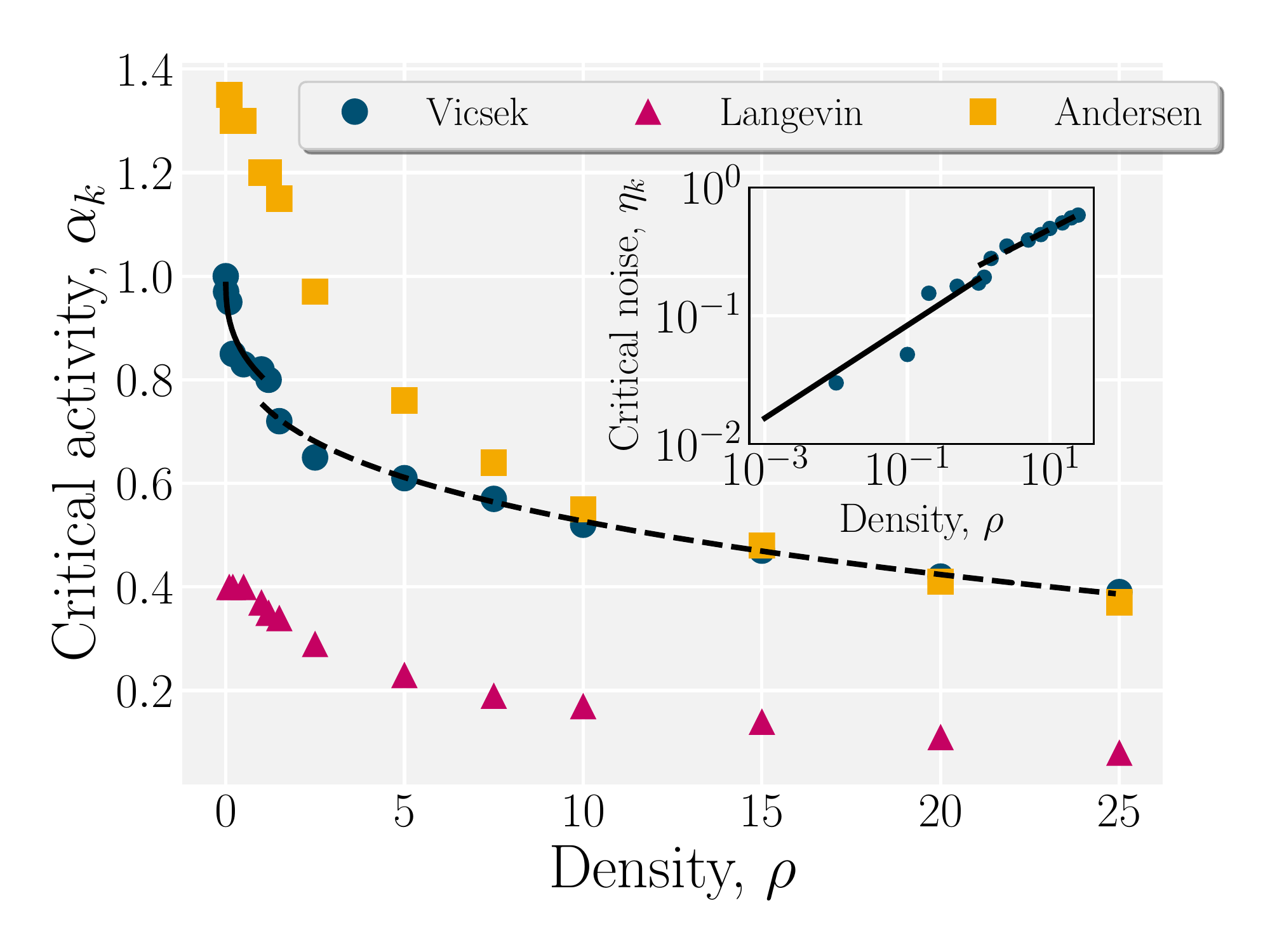}
    \caption{
        \label{fig:ActDens} 
        Critical activity $\alpha_k$ decays as density $\rho$ increases for each of the AP-MPCD collision operators. 
        The solid black lines correspond to the low density systems ($\rho \leq 1$) with scaling $\alpha_k-1\sim\rho^{-1/2}$. 
        The dashed black line corresponds to the high density systems ($\rho > 1$) with scaling $\alpha_k-1\sim\rho^{-1/3}$. 
        The inset shows logarithmic axes of critical noise $\eta_k$ (\eq{eq:activity}) for Vicsek-MPCD.
    }
\end{figure}

\begin{figure}[tb]
    \centering
    \includegraphics[width=0.5\textwidth]{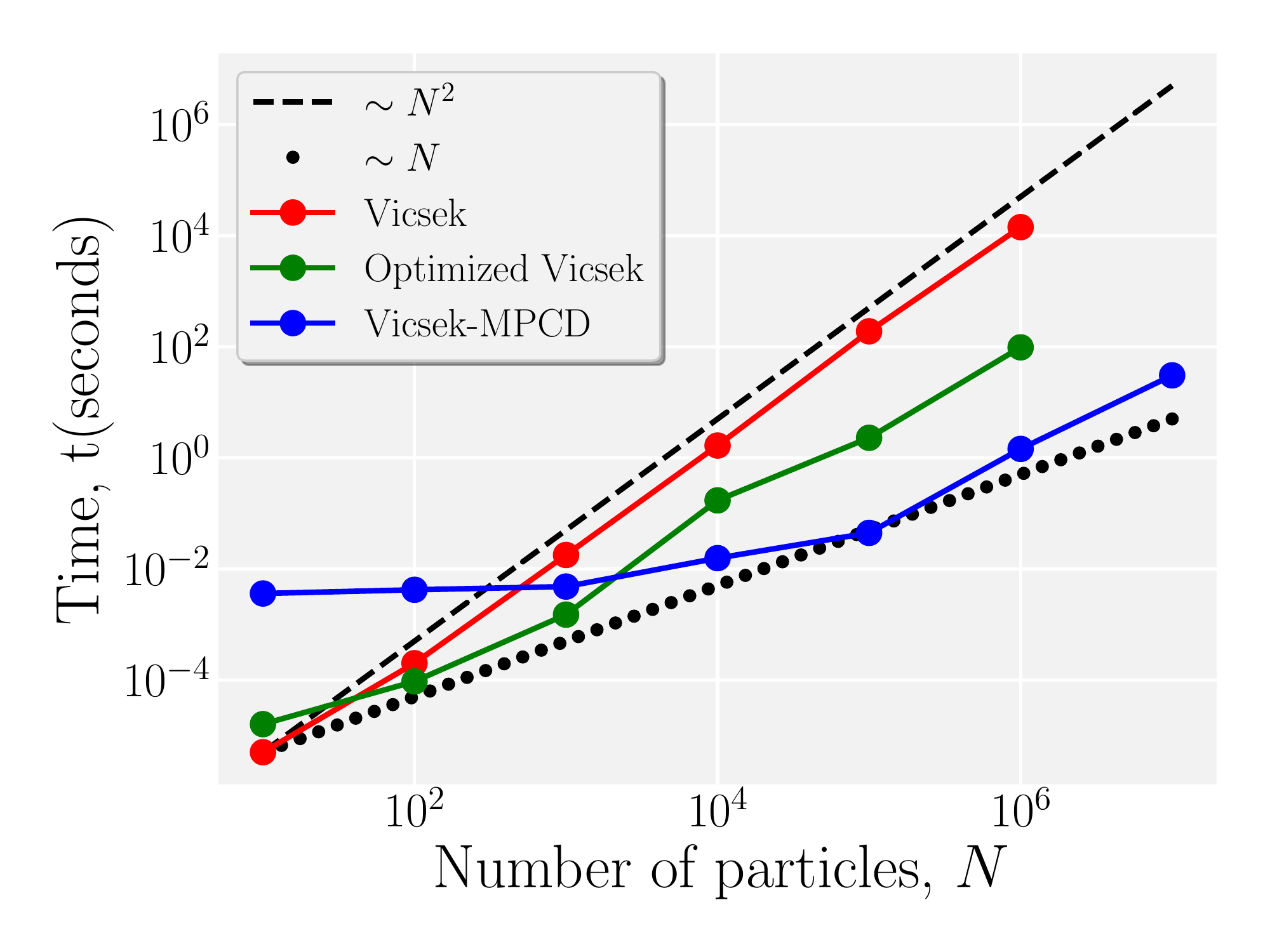}
    \caption{
        \label{fig:modelsPerfomance} 
        \correctText{}{Comparison of algorithms' scaling with number of particles $N$. 
        The Vicsek algorithm without a neighbour list scales quadratically with particle number as $\sim N^2$ (red). 
        Utilising a neighbour lists can improve the scaling to $\sim N\ln(N)$ (green). 
        The hybrid Vicsek-MPCD scales as $\sim N$ for sufficiently large numbers of particles (blue). 
        The black dashed line shows a scaling of $\sim N^2$ and the black dotted line shows $\sim N$ for comparison.}
        }
\end{figure}

For small densities ($\rho \lesssim 2.5$), the critical order parameter decays rapidly but slows as the density increases (\fig{fig:ActDens}; circular markers). 
In both the dilute and dense regimes, the critical activity follows a power law $\alpha_k - 1 = h \rho^n$. 
For $\rho < 1$, $h=0.261\pm0.005$ and the exponent is $n=0.473\pm0.003$, where the uncertainties are the standard deviation between fits for three separate realisations of each parameter. 
This is consistent with the predicted exponent $n=1/D$ in $D=2$ dimensions for low densities ($\rho \leq 1$)~\cite{Chate2008}. 
When the average number of particles per cell exceeds $1$, there is a change in the scaling (\fig{fig:ActDens}). 
This is because particles are continually interacting with neighbours, rather than experiencing discrete pair collisions. 
For larger densities ($\rho>1$), $h=0.238\pm0.008$ and $n=0.296\pm0.026$.
We are not aware of an analytical prediction for the scaling of the critical activity for flocking in the large density limit. 
These suggest the critical activity scales with the density as $\alpha_k \sim \rho^{-1/2}$ in the dilute limit and as $\alpha_k \sim \rho^{-1/3}$ in the dense limit. 

\correctText{}{Since the Vicsek model (\sctn{sctn:vicsek-model}) must calculate the distances between all pairs of particles, the simulation time scales as $\sim N^2$, which can be improved to $\sim N\ln(N)$ using neighbour lists (\fig{fig:modelsPerfomance}). 
On the other hand, the hybrid Vicsek-MPCD need only sort particles into cells within which multi-particle collision operations occur and so scales as $\sim N$ for sufficiently large numbers of particles (\fig{fig:modelsPerfomance}).}

In conclusion, the Vicsek-MPCD algorithm exhibits collective dynamics that are consistent with the traditional Vicsek model.
In the following sections, we will use what we have learned from hybridising MPCD and the Vicsek model to design two novel MPCD algorithms for simulating active polar fluids.

\section{Andersen Active Polar MPCD Algorithm}
\label{sctn:andersen-vicsek}

This section introduces an Andersen active polar MPCD (AAP-MPCD) algorithm that builds on the passive Andersen-thermostatted MPCD algorithm (\sctn{sctn:AndersenMPCD}) and demonstrates that it exhibits a flocking transition similar to the Vicsek model from \sctn{sctn:vicsekMPCD}.

\subsection{Andersen Active Polar MPCD Collision Operator}\label{sctn:Andersen-VicsekCollOp}
\begin{figure}[tb]
    \centering
    \includegraphics[width=0.25\textwidth]{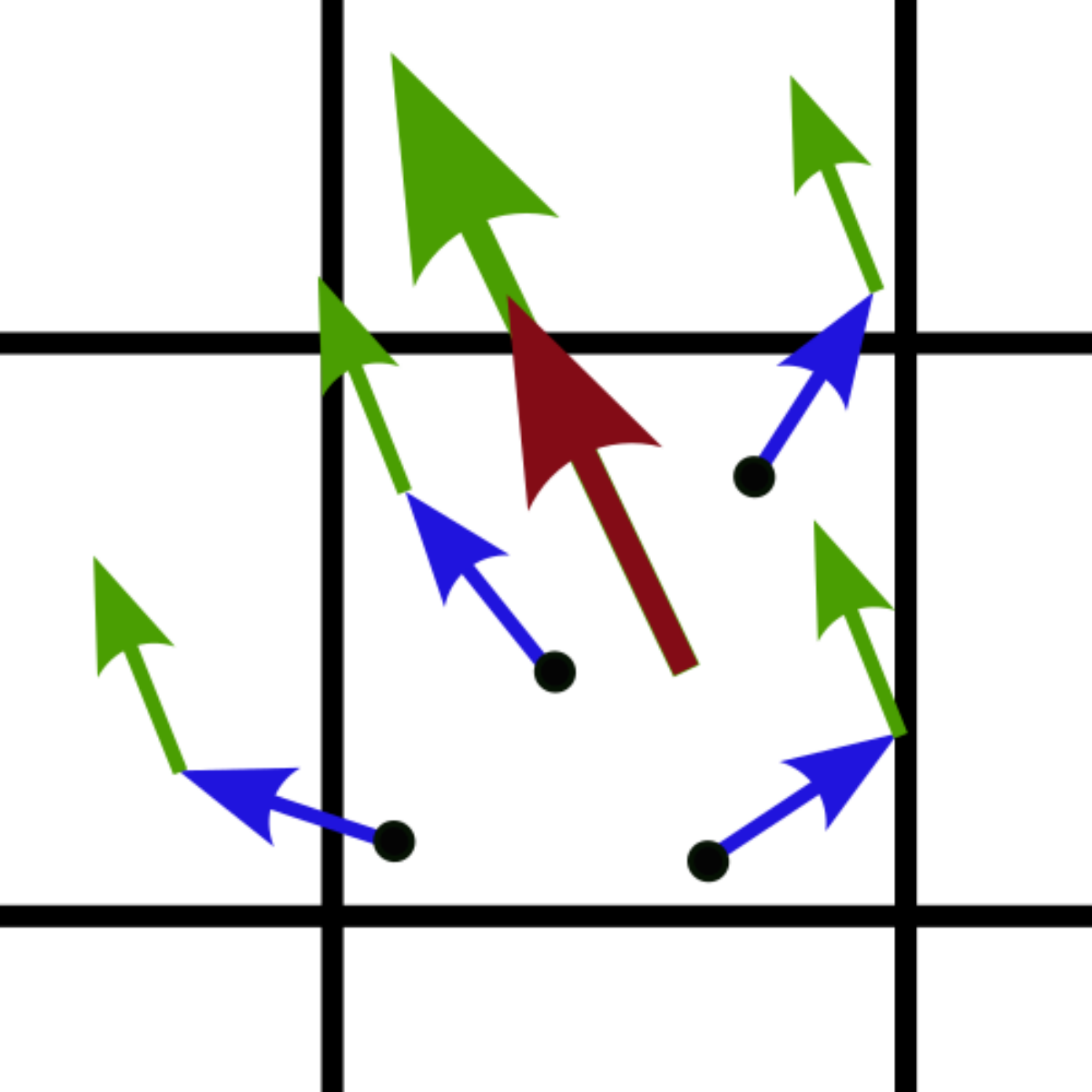}
    \caption{
        \label{fig:VicsekModelAndersenSchemeVicsekPart} 
        Activity in AAP-MPCD (\eq{eq:andersen-vicsekMPCD}) and LAP-MPCD (\eq{eq:Langevin}) drives the velocities of each particle (blue arrows) towards a speed $\alpha$. 
        Green arrow represent the active impulse on each particle and on the centre of mass velocity (red arrow).
    }
\end{figure}

To simulate an active polar fluid with thermostatted particle velocities, the speeds must fluctuate and so the velocities cannot simply be rotated. \correctText{}{Noise enters the AAP-MPCD algorithm as a thermostat that changes both the heading and magnitude of velocity. 
The active term in Eq. 18 causes fluctuations in the speed to relax towards a value of $\alpha$ but, unlike in the Vicsek model, the speed can vary, which is essential for non-dry systems.}
\correctText{Instead,}{} AAP-MPCD sets a target speed $\alpha$ for all the particles and dictates that their velocities approach $\alpha$ in the cell-averaged direction $\hatcm{c} = \vcm{c} / \velcm{c}$ where $\velcm{c} = \abs{\vcm{c}}$. 
The rate at which speeds approach $\alpha$ is controlled by a relaxation parameter $\tau$. In addition, the velocities are subject to random kicks $\vec{\xi}_i$ from a thermal bath.
A collision operator that actively biases the particles towards the desired speed is
\begin{align}
    \vec{R}_{ic}\left(\vcm{c},\Vec{v}_i;t\right) &= \vcm{c}\left(t\right) + \tau \left[ \alpha \hatcm{c}\left(t\right) - \vec{\vel}_i\left(t\right) \right] \nn\\ 
    &\qquad + \vec{\xi}_i - \av{ \vec{\xi} }_c.
    \label{eq:andersen-vicsekMPCD}
\end{align}
This collision operator is identical to Andersen\correctText{-}{ }MPCD collision (\eq{eq:andersen}), except that the additional active term $\tau \left[ \alpha \hatcm{c}\left(t\right) - \vec{\vel}_i\left(t\right) \right]$ drives each velocity towards the target speed $\alpha$ in the average direction of motion $\hatcm{c}$ (\fig{fig:VicsekModelAndersenSchemeVicsekPart}). 

While passive Andersen MPCD conserves momentum, AAP-MPCD breaks momentum conservation, just as the Vicsek model does. Assuming all the particles have the same mass, the local centre of mass velocity after the collision operation is obtained by using \eq{eq:andersen-vicsekMPCD} in \eq{eq:vicsekcollision} and averaging over the cell $c$. 
The centre of mass velocity is updated to
\begin{align}
    \label{eq:aap-MPCD}
    \vcm{c}(t+\dt) &= \av{ \vcm{c}\left(t\right) + \tau \left[ \alpha \hatcm{c}\left(t\right) - \vec{\vel}_i\left(t\right) \right] + \vec{\xi}_i - \av{ \vec{\xi} }_c }_c \nn \\ 
    &= \vcm{c}\left(t\right) + \tau \left[ \alpha - \velcm{c}\left(t\right) \right] \hatcm{c}\left(t\right).
\end{align}
This shows that $\vcm{c}(t+\dt)\neq\vcm{c}(t)$ and momentum is not conserved since the centre of mass velocity changes by $\tau(\alpha - \velcm{c})\hatcm{c}$. 
Only when a cell has a centre of mass speed $\velcm{c}=\alpha$ or $\tau=0$ is the speed unchanged.

AAP-MPCD reduces to the passive momentum-conserving version of Andersen\correctText{-}{ }MPCD (\sctn{sctn:AndersenMPCD}) when $\tau = 0$. So AAP-MPCD seems ``wetter'' for smaller values of $\tau > 0$. 
Additionally, the relaxation has a maximum value of $\tau = 1$, which can be seen by rewriting \eq{eq:aap-MPCD} as $\vcm{c}(t+\dt)=(1-\tau)\vcm{c} + \tau\alpha \hatcm{c}$. 
This form shows the limiting values are $\tau = 0$ and $\tau = 1$. 
In the limit of $\tau = 0$, AAP-MPCD is passive and fully conserves momentum (\textit{i.e.} is ``wet''), whereas when $\tau = 1$ the speed of the MPCD cell immediately becomes $\alpha$ and the system is entirely ``dry''. 
When $0 < \tau < 1$, the fluid is intermediate between wet and dry with $\tau$ controlling the rate of momentum injection.
On the other hand, setting the target speed to zero ($\alpha = 0$) does not remove impulses but rather creates a frictional damping towards a non-flowing state. 

\subsection{AAP-MPCD Results}
Since the AAP-MPCD collision operator has a target speed $\alpha$ rather than the set Vicsek speed $v_0$, the average order parameter (\eq{eq:orderParam}) is redefined as
 \begin{equation}
  \phi = \frac{\abs{\av{\vec{v}}}}{\alpha} = \frac{1}{N \alpha} \abs{\Sigma_{i=1}^{N} \vec{\vel}_i}.
  \label{eq:reDefOrderParam}
\end{equation}
We run simulations with densities $\rho \in [0.001, 25]$ and target speeds in the interval $\alpha \in [0, 2]$. 
The relaxation rate is set to $\tau = 0.5$. 
The initial velocities of the particles are drawn from a uniform isotropic distribution function (\eq{eq:initDist}). 

\begin{figure}[tb]
    \centering
    \includegraphics[width=0.5\textwidth]{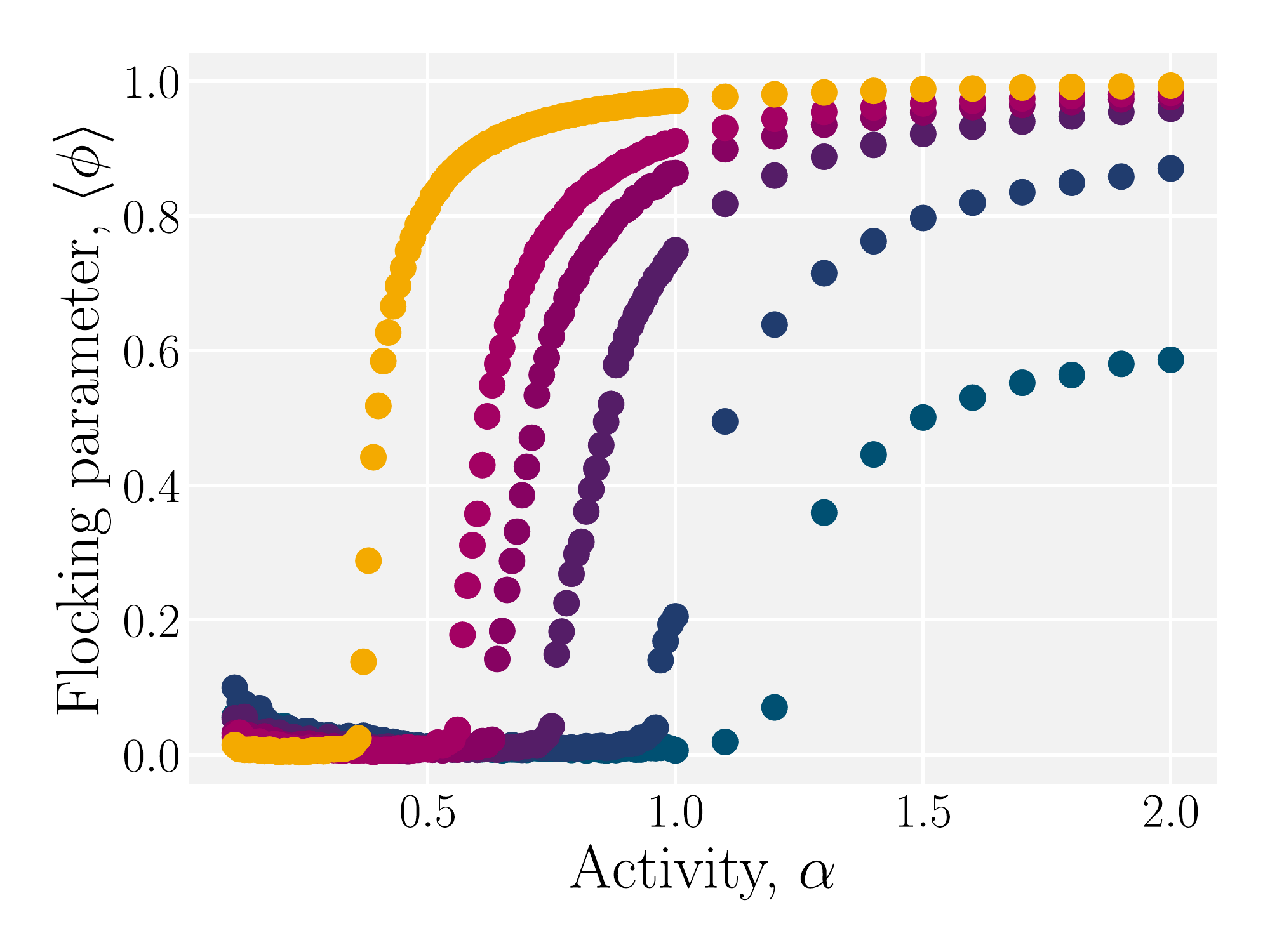}
    \caption{
        \label{fig:VicsekModelAndersen} 
        The flocking transition in AAP\correctText{}{-}MPCD. 
        The time-averaged flocking order parameter $\av{ \phi }$ is the magnitude of the average velocity normalised by activity $\alpha$ (\eq{eq:reDefOrderParam}).
        The colouring represents the density $\rho$ and is the same as in \fig{fig:VicsekModelFigure}. 
    }
\end{figure}

The AAP-MPCD system is found to exhibit a flocking transition (\fig{fig:VicsekModelAndersen}), similar to the Vicsek-MPCD (\fig{fig:VicsekModelFigure}). 
Though AAP-MPCD matches the qualitative behaviour of the Vicsek model, there are quantitative differences.
For instance, the rise of net transport velocity is steeper than observed in the Vicsek-MPCD model (\fig{fig:VicsekModelFigure}). 
After an initial rise, the time-averaged flocking order parameter $\av{ \phi }$ reaches a plateau. 
Small densities are not able to reach their maximum flocking speed of $\alpha$ since $\alpha$ and $\tau$ are not large enough. 
On the other hand, dense systems reach their flocking state even for small $\alpha$ and maintain their order against thermal fluctuations. 

A number of differences exist between the hybrid Vicsek-MPCD and AAP-MPCD algorithms. Most importantly, $\alpha$ represents the competition between active ordering and noise in two very different ways: In the Vicsek-MPCD algorithm $\alpha=1-\eta$ represents the degree of orientational aligning against angular noise, while in AAP-MPCD $\alpha$ is a target speed that should be compared to the thermal speeds $\sqrt{2\kbt/m}$ imposed by the thermostat (\eq{eq:vicsekDist}).
\correctText{}{Unlike the Vicsek model in AAP-MPCD noise enters as a thermostat that changes both the heading and magnitude of velocity, but both algorithms need a source of noise to compete with active alignment this source is a temperature of the system.}
Because of the different activity behaviour in AAP-MPCD, the critical activity also behaves differently. 
Critical activities of AAP-MPCD are larger than the critical values of Vicsek-MPCD for $\tau=0.5$ and $\rho\lesssim10$ but have similar values for $\rho\gtrsim10$ (\fig{fig:ActDens}). 

In summary, particles in AAP-MPCD approach a target speed subject to thermal fluctuations rather than having a fixed speed. 
Yet, AAP-MPCD is seen to exhibit a flocking transition, qualitatively agreeing with the Vicsek model. It does so without being perfectly dry. 

\section{Langevin Active Polar MPCD Algorithm}
\label{sctn:langevin-vicsek}
Building on the Andersen active polar MPCD (\sctn{sctn:andersen-vicsek}), this section introduces an active polar MPCD based on the Langevin\correctText{-}{ }MPCD algorithm. 
The desired properties of the collision operator are the same as in \sctn{sctn:andersen-vicsek}: to have a preferred speed and random thermal forces, but now directly including dissipative terms. 
The results indicate that dissipation does not only preserve the flocking-like behaviour, but shifts the critical activity to lower values. 

\begin{figure}[tb]
    \centering
    \includegraphics[width=0.5\textwidth]{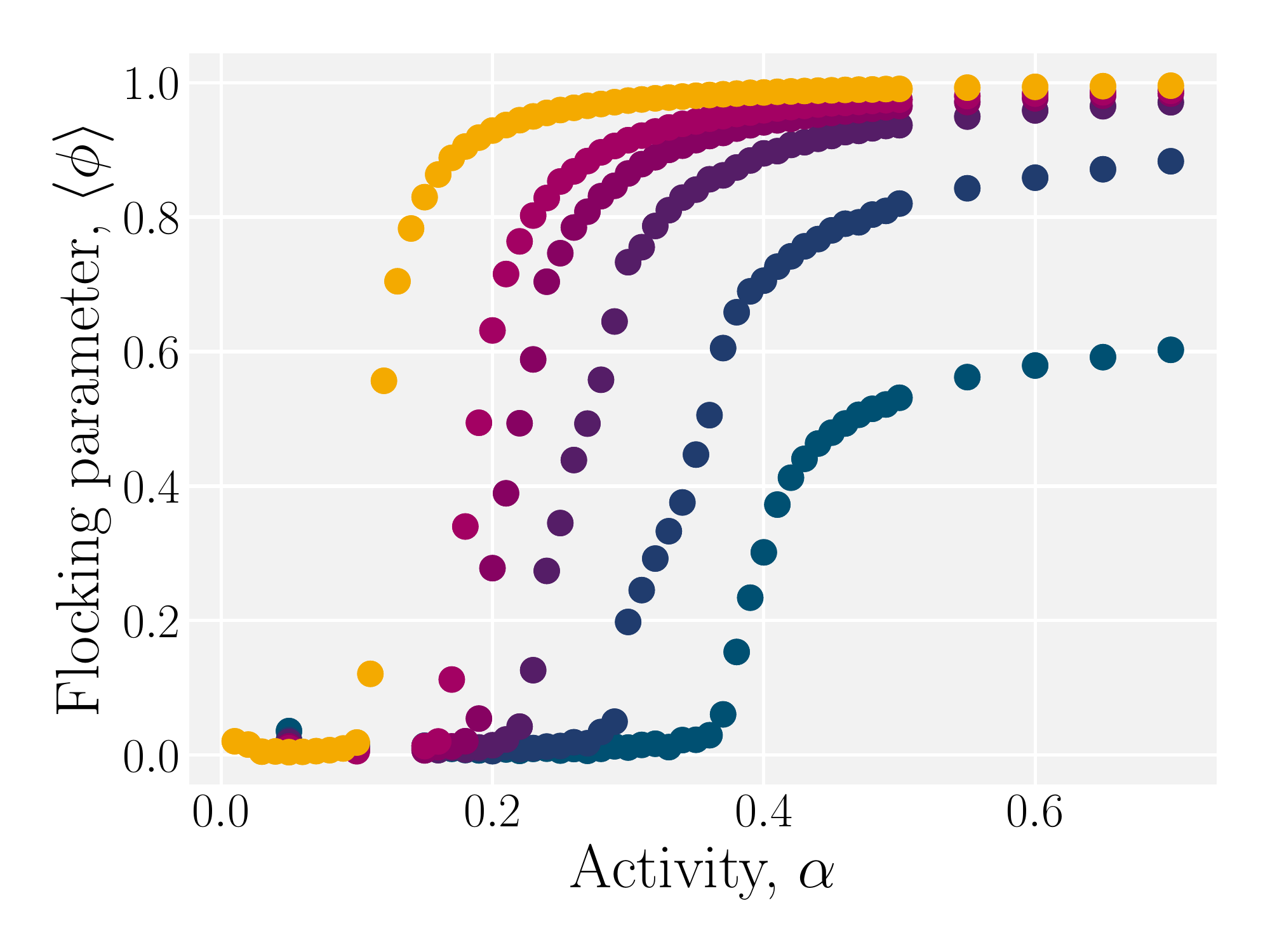}
    \caption{
        \label{fig:VicsekModelLang} 
        The Langevin-AP MPCD flocking transition. 
        The time-averaged flocking order parameter $\av{ \phi }$ is the magnitude of the average velocity normalised by $\alpha$ (\eq{eq:reDefOrderParam}).
        The colouring represents the density $\rho$ and is the same as in \fig{fig:VicsekModelFigure}. 
    }
\end{figure}

\subsection{Langevin Active Polar MPCD Collision Operator}

In the Langevin Active Polar MPCD (LAP-MPCD), the speeds of the particles fluctuate due to the thermal bath. 
The difference with the AAP-MPCD algorithm is that LAP-MPCD should include dissipative force coefficients based on the drag coefficient $\gamma$. 
With this, we propose the LAP-MPCD collision operator
\begin{align}
  \vec{R}_{ic}\left(\vcm{c},\Vec{v}_i;t\right) &= \vcm{c}\left(t\right) + \tau \left[ \alpha \hatcm{c}\left(t\right)-\vec{\vel}_i\left(t\right) \right] \nn\\ 
  &\qquad + b\left[ \vec{\vel}_i\left(t\right) - \vcm{c}\left(t\right) \right] \nn\\ 
  &\qquad\qquad + \ell\left( \vec{\xi}_i - \av{ \vec{\xi} }_c \right),
  \label{eq:Langevin}
\end{align}
in which the last two terms come from the passive Langevin algorithm (\eq{eq:LangevinPassive}) with $b$ and $\ell$ given by \eq{eq:b} and \eq{eq:ell}. 
The second term is the same active contribution as in AAP-MPCD (\eq{eq:andersen-vicsekMPCD}), with $\alpha$ and $\tau$ playing the same roles of target speed and relaxation parameter respectively. 
As in the Andersen case, the centre of mass velocity of cell $c$ after the collision operation obeys \eq{eq:aap-MPCD}:
the LAP-MPCD algorithm also does not preserve momentum.

\subsection{LAP-MPCD Results}
Simulations are run with densities \correctText{$\rho \in [0.08, 25]$}{$\rho \in [0.001, 25]$} and activity in the interval $\alpha \in [0, 1]$. 
The initial velocities of the particles once again are drawn from a uniform isotropic distribution (\eq{eq:initDist}). 
The relaxation parameter is set to $\tau = 0.5$ and the friction coefficient to $\gamma = 1$. 
The model exhibits the same plateau as the other active polar models, with the average flocking order parameter $\av{ \phi }$ rising even more steeply than or the hybrid Vicsek-MPCD or AAP-MPCD (\fig{fig:VicsekModelLang}). 
The LAP-MPCD with $\gamma=1$ transitions to flocking at activities $\alpha_k$ that are approximately three times smaller than the AAP-MPCD model (\fig{fig:ActDens}). 

To understand why LAP-MPCD has a lower critical activity than AAP-MPCD, consider that the transition to an ordered flocking state occurs when the distribution of \correctText{velocity}{the velocities} narrows. 
In AAP-MPCD (\eq{eq:andersen-vicsekMPCD}), it is the active term $\tau \left[ \alpha \hatcm{c} - \vec{\vel}_i \right]$ alone that narrows the distribution. 
However, in LAP-MPCD, the friction term $b\left[ \vec{\vel}_i - \vcm{c} \right]$ also acts to narrow the distribution about the centre of mass velocity. 
Thus, the friction term aids activity in aligning the velocities and results in a lower critical activity $\alpha_k$.

All three active polar MPCD models show a flocking transition. 
The next sections consider perturbations to the flocking state through external fields or complex geometries. 
We choose to focus on LAP-MPCD since AAP-MPCD can be viewed as a specific case of LAP-MPCD (\sctn{sctn:Langevin-MPCD}). 

\begin{figure}[tb]
    \centering
    \includegraphics[width=0.5\textwidth]{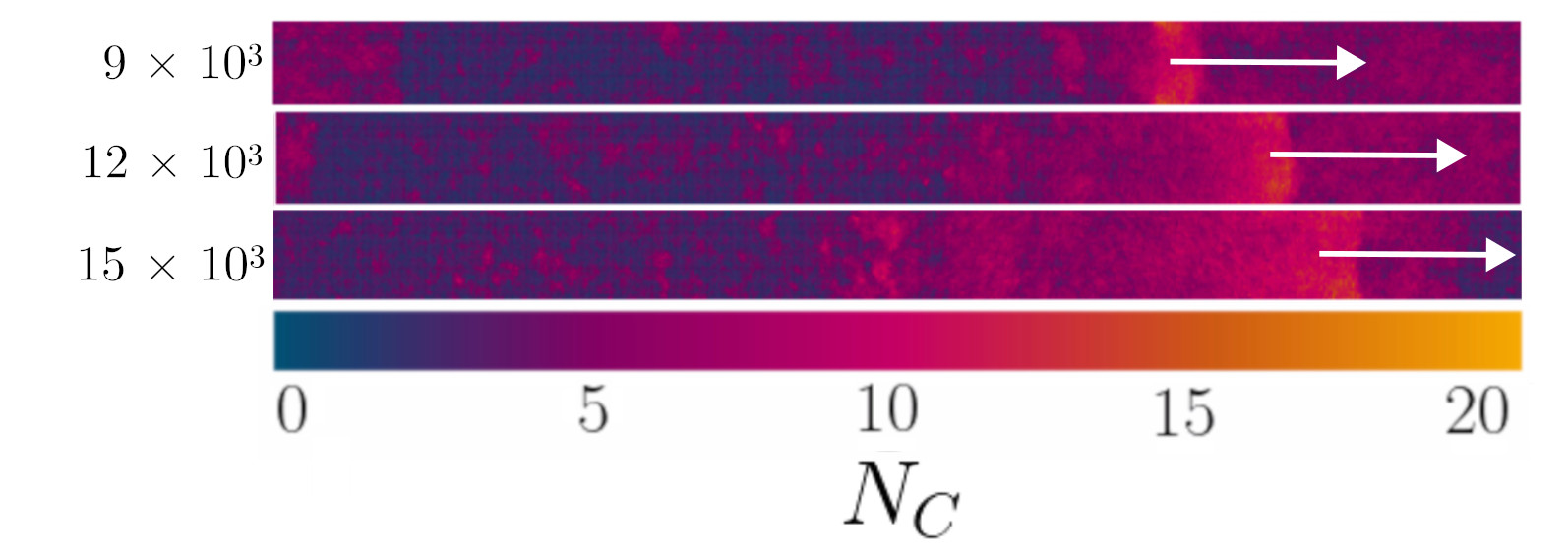}
    \caption{
        \label{fig:BandsExample} 
        Snapshots of a travelling density band in a $1200 \times 75$ channel. 
        The  global density is $\rho = 10$, the activity is $\alpha = 0.22$ and the colour bar shows the number of particle $N_c$ in each MPCD cell.
        The time step is labelled on the left and the difference in time between snapshots is $3 \times 10^3$.
        White arrows represents the direction of motion of the band.
    }
\end{figure}

\subsection{Channel-Confined Flocking}
\label{sctn:banding}

One of the unexpected aspects of the Vicsek model is the formation of bands in channels with periodic boundary conditions~\cite{Aldana2009}. 
Although it was originally thought that the Vicsek model possesses a second-order phase transition to the flocking state~\cite{vicsekspp}, more detailed studies demonstrated that it is a weak first-order transition with a tight range of coexistence~\cite{BinderVirnau}, which manifests as high density travelling bands moving through low density surroundings. 
We have already seen a suggestion of this in the negative dip of the Binder cumulants (\fig{fig:BinderCumulant}), but we now explicitly study the traveling bands. 
To facilitate the formation of bands, the system is set to a periodic channel geometry. 
The system is rectangular with size $\correctMath{}{L_x \times L_y =}1200 \times 75$. 
The boundary conditions are periodic in all directions and the global particle density is $\rho=10$. 
There is no warmup time ($\correctMath{T}{\Tint}_W=0$) because we are interested in the process of bands evolution. 

\begin{figure}[tb]
    \centering
    \includegraphics[width=0.5\textwidth]{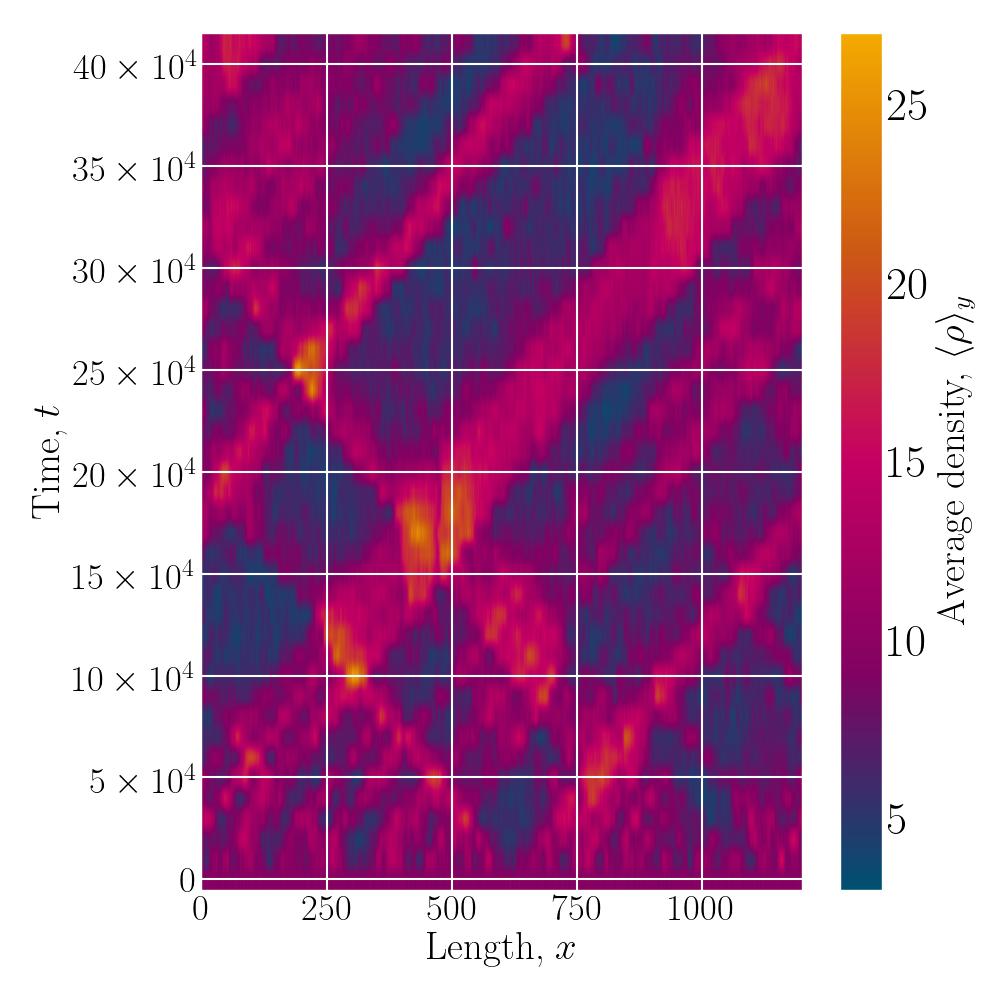}
    \caption{
        \label{fig:SpaceTimePlotAct0} 
        Kymograph of the channel-averaged density $\av{\rho}_y$ of LAP-MPCD in a rectangular channel with periodic boundary conditions for global particles density $\rho=10$ and activity $\alpha=0.22$. 
    }
\end{figure}

In the vicinity of the phase transition point, LAP-MPCD exhibits coexistence via the formation of travelling density bands (\fig{fig:BandsExample} \correctText{}{and \movie{mov:bands}}). 
The bands consist of high density regions that travel coherently along the axis of the channel. 
Since the bands travel along the channel in the $\pm\hat{\vec{x}}$-direction, the density is averaged across the channel $\av{\rho}_y(x;t) = L_y^{-1} \sum_{y=0}^{L_y} N_c(x,y;t)$ and plotted as a kymograph or space-time diagram (\fig{fig:SpaceTimePlotAct0}). 
The kymographs show that bands form and travel through the system. 
At early times, bands are equally likely to move in the $+\hat{\vec{x}}$-direction as the $-\hat{\vec{x}}$-direction and many narrow bands exist, criss-crossing the system (\fig{fig:SpaceTimePlotAct0}; $t \lesssim 15\times10^4$). 
However, as time progresses many of the bands merge and eventually the one direction is spontaneously selected as the average direction of flocking (the $+\hat{\vec{x}}$-direction for the particular realisation shown in in \fig{fig:SpaceTimePlotAct0}). 
By the end of the simulation, all the bands are travelling in the same direction (\fig{fig:SpaceTimePlotAct0}). 
The slopes in the kymographs are linear, observed to be the same for each band and constant in time (\fig{fig:SpaceTimePlotAct0}). 
The group speed can be measured as the slope of the wave front of high density lines in the kymographs. 

\begin{figure}[tb]
    \centering
    \includegraphics[width=0.5\textwidth]{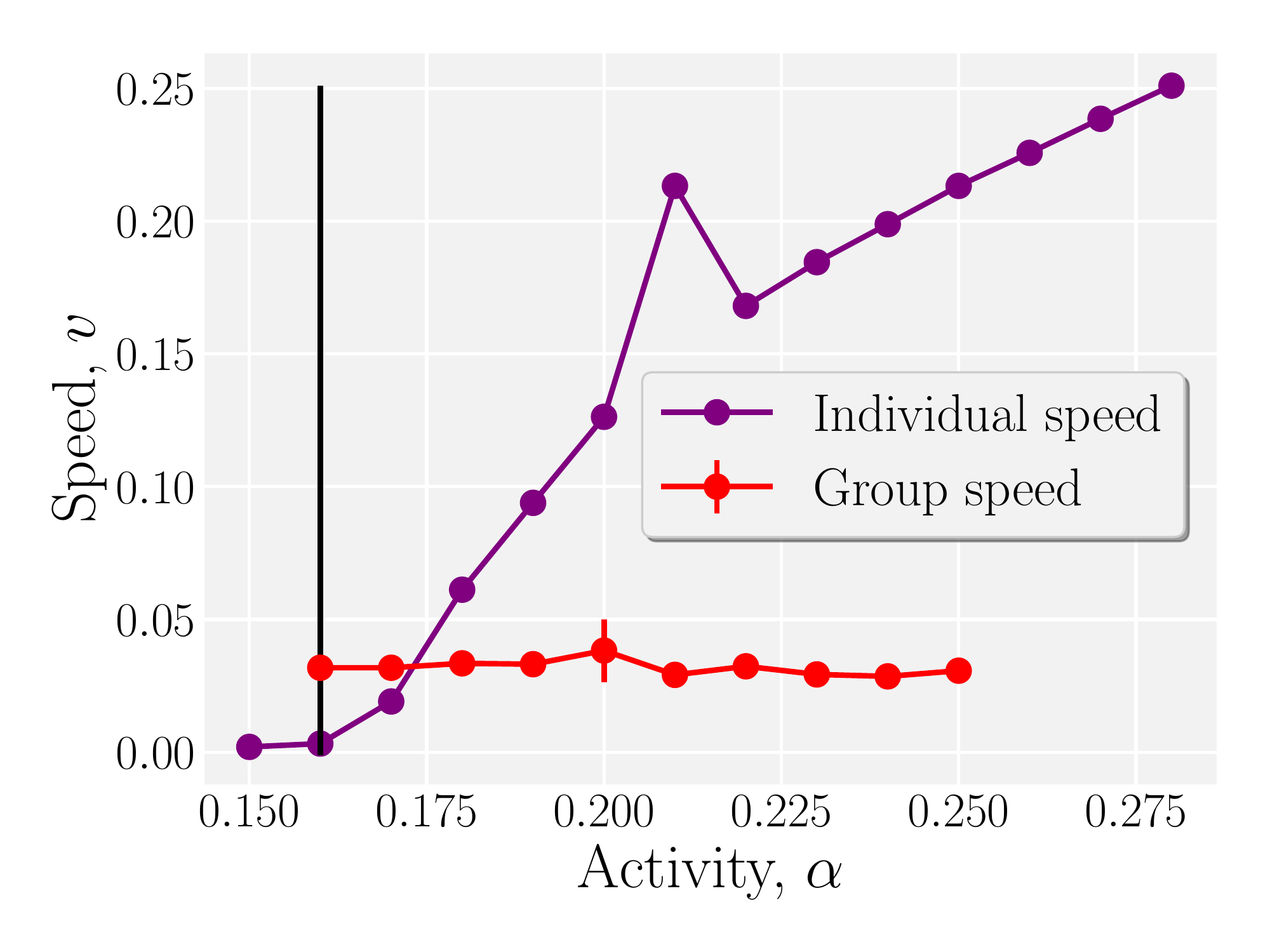}
    \caption{
        \label{fig:ParticleSpeeds} 
        Comparison of group speed of high density bands to individual speeds just above the critical activity $\alpha_k=0.16$ (vertical black line). 
        Purple markers show the average speed of individual particles $\av{\left|\vec{v}\right|}$. 
        Red markers show group speed of the band measured from the kymographs (\fig{fig:SpaceTimePlotAct0}). \correctText{}{Error bars represent standard deviation between bands.}
        The global density is $\rho=10$, and the system size is $1200 \times 75$.
        Measurements taken after $\correctMath{T}{\Tint} >15 \times 10^4$.
    }
\end{figure}

We find coexistence in the form of travelling bands in the activity range $0.16 \le \alpha\le 0.25$ in the case of global density $\rho = 10$. 
While individual particles speed up as activity $\alpha$ increases, the group speed of the bands remains constant with activity in this regime (\fig{fig:ParticleSpeeds}). 
For a small region of parameter space above the critical activity $\alpha_k$, the group speed is slightly faster than the individual speeds $\av{\abs{\vec{v}}}$. 
However, by $\alpha=0.18$, the  individual speed is greater than the group speed. 
Although the group speed is constant with activity immediately above the critical point, the system-averaged velocity $\phi \propto \abs{\av{\vec{v}}}$ increases rapidly in this regime (\fig{fig:VicsekModelLang}). 
This implies that the near-transition increase in $\phi$ occurs not due to acceleration of the bands but rather because the fraction of particles in the coherently moving dense bands increases with activity. 
\correctText{}{The group speed is not found to be impacted by finite size effects since larger systems ($L_y>1200$) exhibit comparable group speeds.} 

In summary, LAP-MPCD reproduces coexistence of disorder and dense coherent travelling density bands for a narrow region of parameter space, which is consistent with the Vicsek model as it under goes a first order transition to the flocking state \cite{Ginelli2008}.

\subsection{Influence of \correctTitle{the}{} Gravity}
\label{sctn:gravity}
While the previous section found that density bands exist for LAP-MPCD, this sections studies how stable these bands are if they are subjected to external perturbation.
We consider the case of gravity $\vec{g}$ changing the velocity of particles each time step according to
\begin{equation}
    \vec{\vel}_i\left(t+\dt\right) = \vec{R}_{ic}\left(\vcm{c},\Vec{v}_i;t\right) + \vec{g} \dt ,
\end{equation}
in which $\vec{R}_{ic}$ is the LAP-MPCD collision operator (\eq{eq:Langevin}). 
The gravitational field is applied in the $-\hat{\vec{y}}$-direction, $\vec{g}=-g\hat{\vec{y}}$. 
For this section, we consider a rectangular system of size $75 \times \correctMath{1200}{8000}$. 
This choice means gravity acts parallel to the long axis of the channel, which is the direction of motion of the density bands. 
The particle density is set to $\rho = 10$ and activity $\alpha=0.22$ is chosen so that the system is in the vicinity of transition point (\fig{fig:ParticleSpeeds}).  
Gravity magnitudes are in the range $g \in \left[10^{-6},10^{-1}\right]$. 

\begin{figure}[tb]
    \centering
    \includegraphics[width=0.4\textwidth]{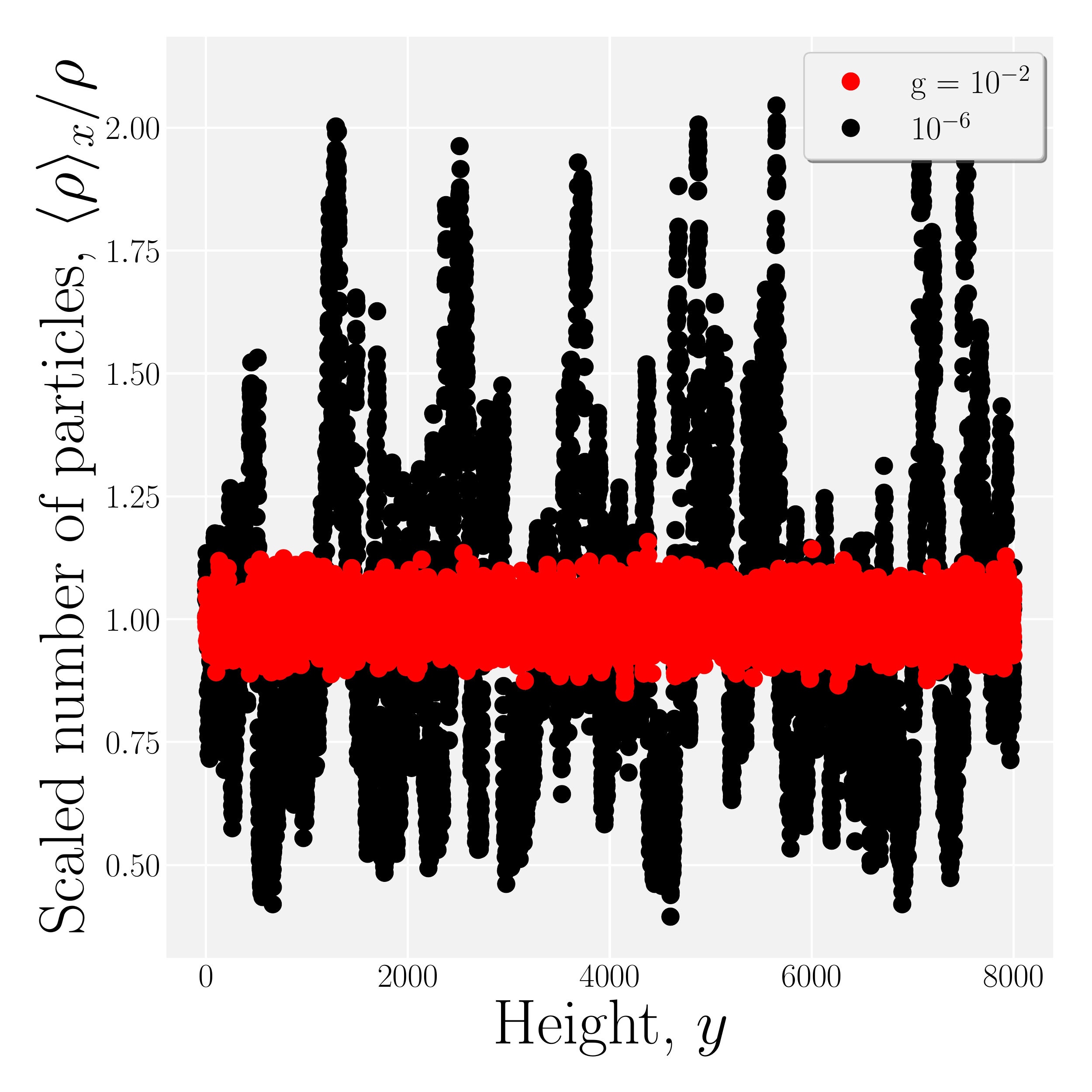}
    \caption{
        \label{fig:GravCorr} 
        Instantaneous number density of particles averaged across the channel $\av{\rho}_x$ and scaled by the global density $\rho$ for gravities $g = 10^{-6}$ (black) and $g = 10^{-2}$ (red). Density is \correctText{$\rho=25$}{$\rho=10$} and activity is $\alpha=0.22$.
    }
\end{figure}

\begin{figure}[tb]
    \centering
    \includegraphics[width=0.5\textwidth]{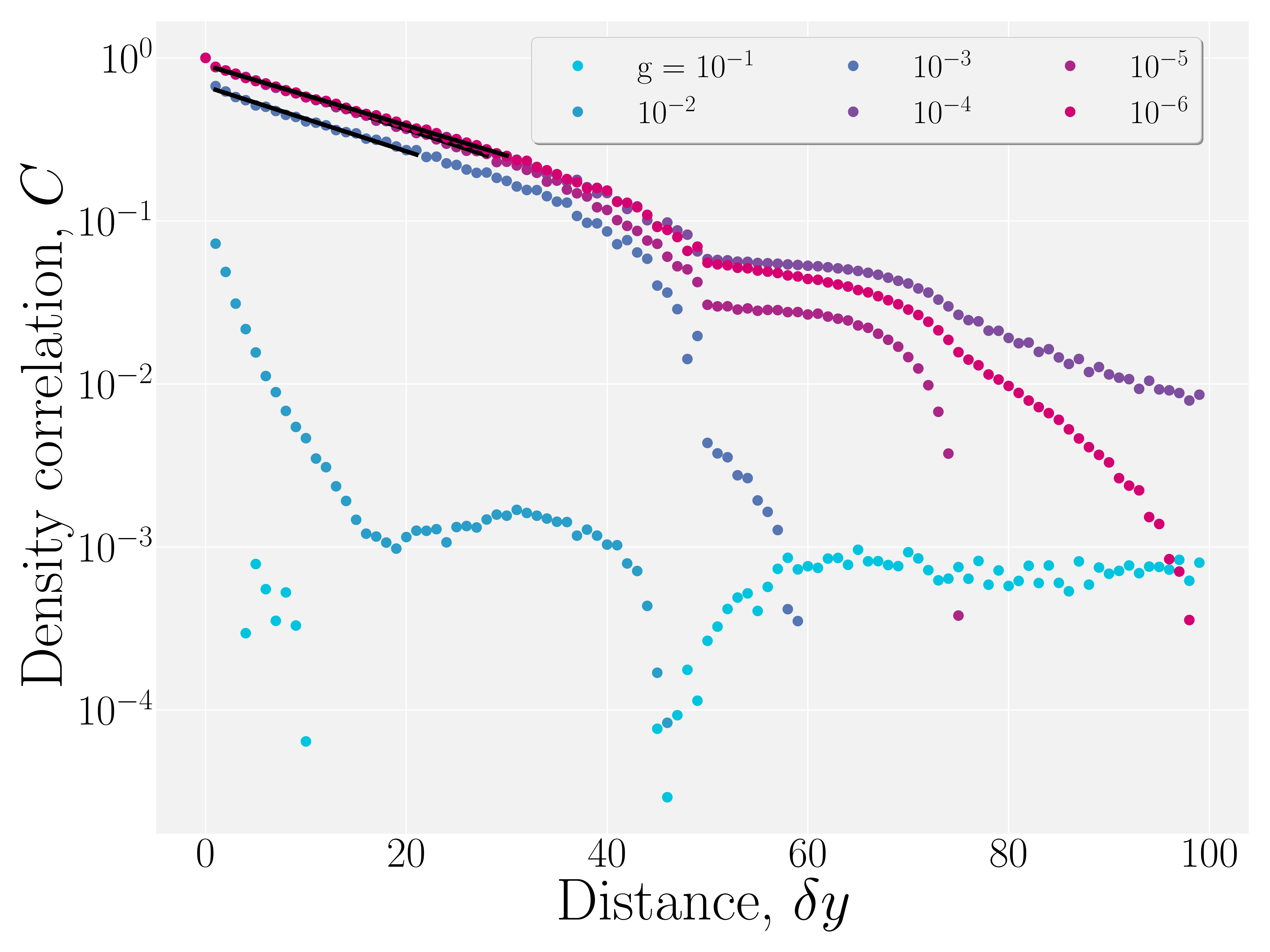}
    \caption{
        \label{fig:Correlation} 
        Spatial autocorrelation $C$ of the number density $\av{\rho}_x$ (\eq{eq:CorrelationFunction}). 
        Solid lines are fits to an exponential decay (\eq{eq:exp}). 
        The system has global density $\rho=10$, activity $\alpha=0.22$ and size $75 \times \correctMath{1200}{8000}$.
    }
\end{figure}

When the gravity is zero or weak, the bands remain stable (\fig{fig:GravCorr}; black), as seen as localised peaks in the density that coherently travel through the system. 
Qualitatively, the peak width is seen to be many tens of MPCD cells wide, in agreement with the zero gravity limit (\fig{fig:SpaceTimePlotAct0}). 
However, as the strength of gravity increases, the bands disappear. 
At $g=10^{-2}$, the density $\av{\rho}_x\left(y\right)$ does not vary substantially about the mean and any deviations from $\rho$ rapidly decay (\fig{fig:GravCorr}; red). 
To quantify the variation about the mean, the spatial autocorrelation function
\begin{align}
    \label{eq:CorrelationFunction}
    C\left(\delta y\right) &= \frac{\av{ \av{\rho}_x\left(y\right) \av{\rho}_x\left(y+\delta y\right) } - \av{ \av{\rho}_x }^2 }{ \av{ \av{\rho}_x^2 } - \av{ \av{\rho}_x }^2 }
\end{align}
is measured, in which $\av{\cdot}_x$ denotes averaging over cells across the channel and $\av{\cdot}$ denotes averaging along the channel and over time.
The autocorrelation function decreases and is fit to an exponential decay
\begin{align}
    \label{eq:exp}
    C\left(\delta y\right) &= C_0 e^{-\delta y/d}
\end{align}
over the range for which $C>1/e$ (\fig{fig:Correlation}). 
The decorrelation length $d$ is interpreted as the band width. 

\begin{figure}[tb]
    \centering
    \includegraphics[width=0.5\textwidth]{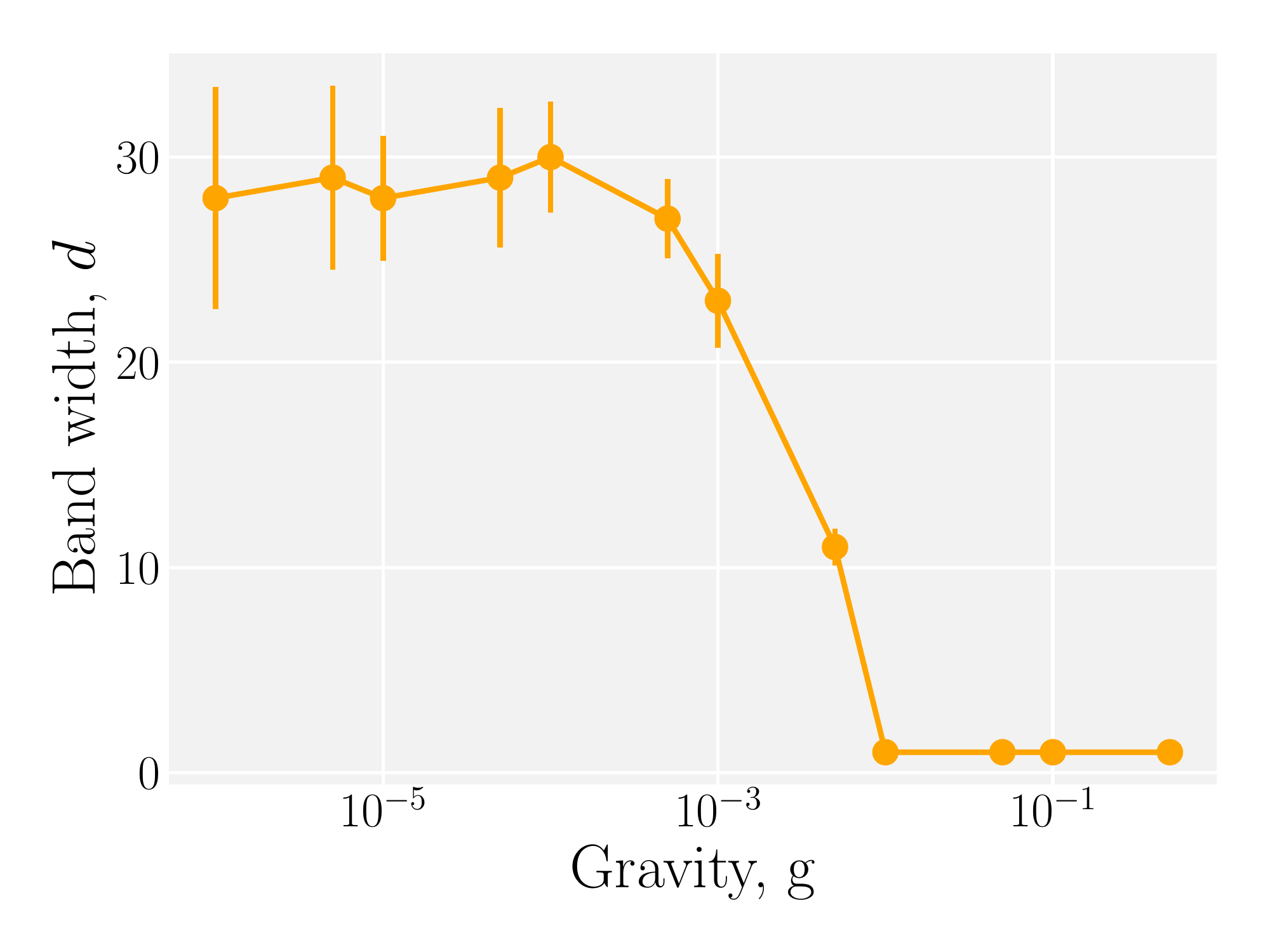}
    \caption{
        \label{fig:BandsDens} 
        Width $d$ of the travelling density bands subject to gravity $g$ \correctText{for various global densities}{for global density $\rho=10$}. 
        The system has activity $\alpha=0.22$ and size $75 \times \correctMath{1200}{8000}$. \correctText{}{Error bars represent the uncertainty on the fit from \fig{fig:Correlation}.}
    }
\end{figure}

When gravity is weak, variations in the density are correlated over many MPCD cells (\fig{fig:Correlation}), which is consistent with \fig{fig:GravCorr} (black). 
For $\rho=10$, the band width $d$ decreases as the strength of gravity increases (\fig{fig:BandsDens}). 
Once $g$ increases above \correctText{$10^{-4}$}{$10^{-2}$}, the width goes to zero (\fig{fig:BandsDens}) since autocorrelation function is an immediate, delta-function drop to zero (\fig{fig:Correlation}). 
This indicates that the system can no longer sustain travelling density bands of finite width --- gravity destroys the bands completely. 
\correctText{Other global densities $\rho$ in the vicinity of the critical point show a similar trend.}{} 
\correctText{For densities above the flocking transition ($\rho \geq 10$) as}{As} gravity increases, the band width $d$ shrinks until $g$ reaches a critical value, at which point $d$ drops to zero signalling that gravity has destroyed the bands (\fig{fig:BandsDens}). 
\correctText{For densities below the flocking transition ($\rho<10$), no bands exist for any $g$.}{}
These results indicate that bands are a relatively fragile state that can be disrupted by a sufficiently strong external forcing. 

\subsection{Asymmetric Obstacles in the MPCD Framework}
\label{sctn:obstacles}

If an external field, such as gravity, can alter the nature of the flocking, can other biases change the dynamics? 
To explore this, we consider flocking in a system of asymmetric obstacles. 
Asymmetric obstacles have previously been demonstrated to act as ratchets, biasing the flocking direction, which can be exploited to direct the motion of self-propelled particles~\cite{Martinez2020}. 

The MPCD framework facilitates the addition of a wide variety of boundaries as surfaces
\begin{equation}
    \label{eq:surface}
     \mathcal{S}_b\left(\vec{r}^*\right) = 0 ,
\end{equation}
in which $b$ is the index labelling the surface and $\vec{r}^*$ is a point on the surface. 
An MPCD particle at position $\vec{r}_i$ is outside boundary $b$ so long as $\mathcal{S}_b\left(\vec{r}_i\right) > 0$ but has impinged on the surface if $\mathcal{S}_b(\vec{r}_i) \leq 0$. 
This causes the impinging particle to be ray-traced back to the surface boundary at position $\vec{r}^*_i$ at time $t^* < t+\dt$, for which the particle path and the surface $\mathcal{S}_b(\vec{r}_i) = 0$ intersect. 
At this point, the MPCD particle collides with the obstacle. 
The velocity of the particle is reflected.
Having collided with the obstacle, the particle may move along its new velocity for the remainder of the time step, $\dt - t^*$.

It remains to define the desired surface (\eq{eq:surface}). 
Let $\vec{q}_b$ be the position of the $b$\textsuperscript{th} obstacle and $\delta \vec{r} = \vec{r}_i-\vec{q}_b \equiv \delta x\hat{\vec{x}} + \delta y\hat{\vec{y}}$ be the position of the $i$\textsuperscript{th} MPCD particle relative to the obstacle with separation $\delta r = \abs{\delta \vec{r}}$. 
One set of surfaces $S_b(\vec{r})$ could be planes, circles or ellipses, which are all members of the class of super-ellipses, for which the surface has the generic form
\begin{equation}
    \label{eq:superellipse}
    \mathcal{S}_b = \left( \frac{\delta x}{A} \right)^{p} + \left( \frac{\delta y}{B} \right)^{p} - \mathcal{R}^{p} . 
\end{equation}
For $p=1$ this is the equation of a plane, for $p=2$ this is an ellipse, and for $p>2$ and even this is a superellipse with four-fold symmetry. 

To break the symmetry, we use an extension to \eq{eq:superellipse} with a more generic form that allows for approximations of more complex shapes, such as triangles, hexagons and stars~\cite{Gielis2003}. 
In 2D, the relative position defines an angle $\varphi=\tan^{-1}\left( \delta x/\delta y \right)$, which can be used to define the surface
\begin{equation}
    \label{eq:Gielis}
    \mathcal{S}_b = \left( \frac{\delta x}{A\cos\left(M\varphi/4\right)} \right)^{p} + \left( \frac{\delta y}{B\sin\left(M\varphi/4\right)} \right)^{p} - \mathcal{R}^{p} . 
\end{equation}
Here, $M\neq4$ breaks the four-fold symmetry, with $M=3$ producing approximate triangles. 

\begin{figure}[tb]
    \centering
    \includegraphics[width=0.5\textwidth]{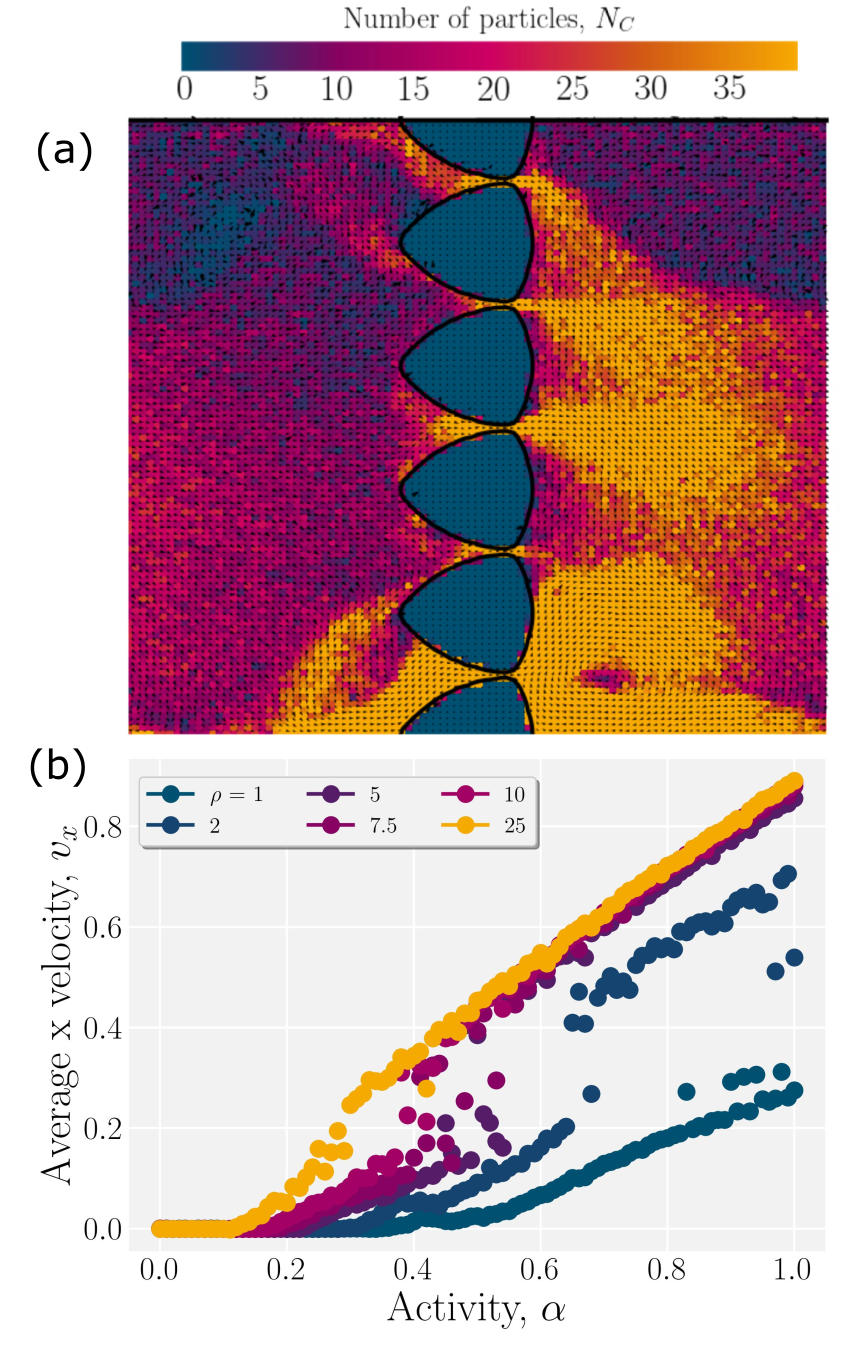}
    \caption{
        \label{fig:VicsekTriangleBarrier} 
        Rectification of flocking by an array of triangular obstacles of size $\mathcal{R}=8$ with gaps of size of $5$.
        (a) Snapshot of a flock moving through funnels showing the velocity (arrows) and density (colour map) fields. 
        \correctText{}{The global density is $\rho=10$ and the activity is $\alpha=0.5$}
        Impermeable walls are located at the top/bottom of the system, while periodic boundaries are at the left/right. 
        (b) The average $\hat{\vec{x}}$-component of the velocity, $v_x$. 
        The funnels cause $v_x>0$ in all cases, indicating rectified motion from left to right.
    }
\end{figure}

\begin{figure}[tb]
    \centering
    \includegraphics[width=0.5\textwidth]{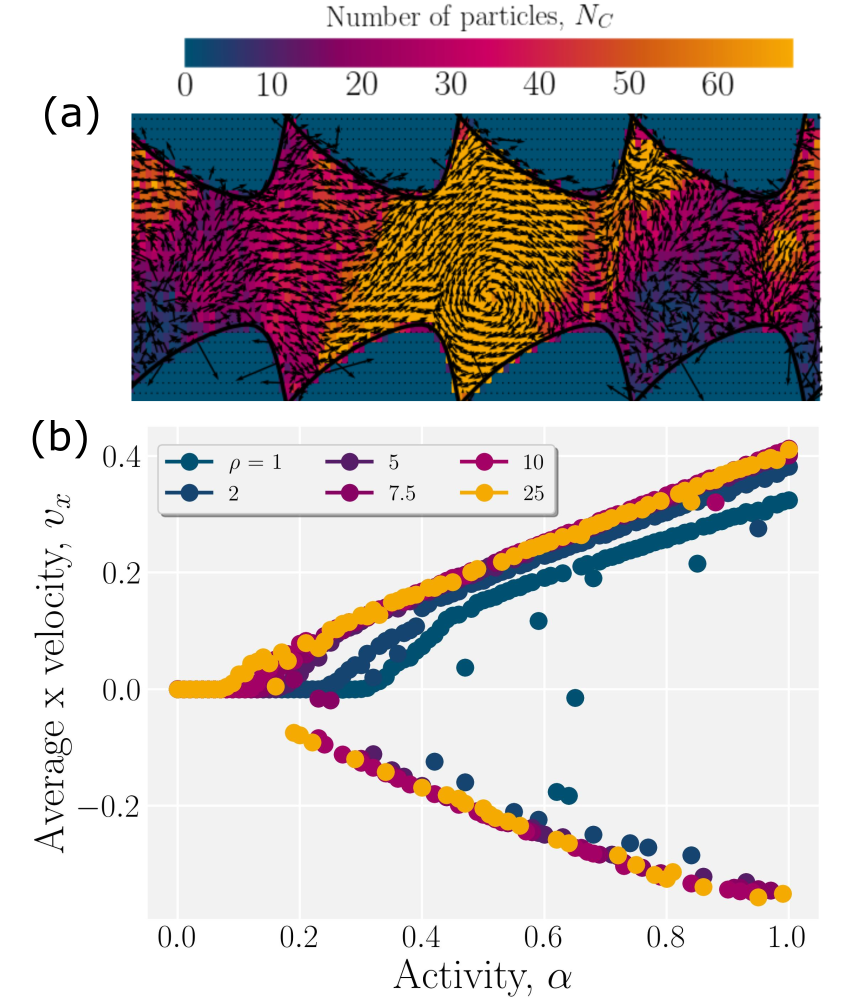}
    \caption{
        \label{fig:VicsekTrianglesChannel} 
        Rectification of flocking by a ratchet channel composed of triangular obstacles. 
        (a) Snapshot of a flock moving through the channel showing the velocity (arrows) and density (colour map) fields. 
        \correctText{}{The global density is $\rho=10$ and the activity is $\alpha=0.5$}
        (b) The average $\hat{\vec{x}}$-component of the velocity, $v_x$. 
        While the majority of collective flocks move in the ``easy direction'' (left to right; $v_x>0$), a minority ($\approx26\%$) move in the opposite direction ($v_x<0$). 
        Periodic boundaries are applied on the left/right ends.
    }
\end{figure}

We simulate two different configurations of triangles as obstacles for the LAP-MPCD model. 
First, we consider an array of triangular obstacles with gaps forming a funnel (\fig{fig:VicsekTriangleBarrier}) and second, we study a channel with ratchet walls (\fig{fig:VicsekTrianglesChannel}). 
In both cases, the global density is $\rho=25$, and all obstacles are triangles with $M=3$, power $p=12$ and $A=B=1$. 
Phantom particles are included in any cell that is intersected by a \correctText{}{triangular obstacle} boundary surface~\cite{Lamura}. 
\correctText{}{Because the MPCD viscosity depends on the particle number density~\cite{Noguchi2008}, cells with partial excluded volume have a systematically lower viscosity~\cite{Lamura}. 
Phantom particles are used to ensure that the viscosity near the boundaries is not artificially lower than the bulk, which is necessary to ensure no-slip boundary conditions. 
The phantom particle velocities are drawn from zero-mean Gaussian distributions, which weigh each cell's centre of mass velocity towards zero.}

Firstly, we consider a barrier of five triangles in a square system of size $L = 100$ (\fig{fig:VicsekTriangleBarrier}(a) \correctText{}{and \movie{mov:funnels}}). 
On the top and the bottom, impermeable \correctText{}{perfect-slip} walls form a wide channel with periodic boundary conditions at the ends, which ensures that flocking must occur in the $\pm\hat{\vec{x}}$ directions. 
The five obstacles are fixed in place at
$\vec{q}_b=50\hat{\vec{x}} + b 20 \hat{\vec{y}}$ \correctText{}{and have no-slip boundaries with phantom particles}. 
Since the size of the triangles is $\mathcal{R}=8$, this leaves a gap of $5$ MPCD cells at the narrowest point between each triangle. 
On the right-hand side, the triangular bases act as a wall with small pores; whereas, on the left-hand side, pairs of triangles act as funnels. 
This rectifies the flocking direction (\fig{fig:VicsekTriangleBarrier}(b)), biasing the flow to move from left to right, completely suppressing the reverse flow. 
By the end of every simulation, the active particles are all flocking in the $+\hat{\vec{x}}$-direction. 
While the funnels rectify the direction of flocking, they also disrupt the phase transition. 
While the flocking transition in bulk exhibited a sharp rise in the flocking order parameter (\fig{fig:VicsekModelLang}), the average $\hat{x}$-component of the velocity increases smoothly in the presence of the asymmetric obstacles (\fig{fig:VicsekTriangleBarrier}(b)). 

The second configuration we consider is a ratchet channel (\fig{fig:VicsekTrianglesChannel}(a) \correctText{}{and \movie{mov:ratchetChannel}}). 
The system is a long channel with size $120 \times 60$. 
Again, impermeable perfect-slip walls on the top and bottom form a channel with periodic boundary conditions at the ends. 
Two rows of \correctText{}{no-slip} triangles with four on each row along the bottom $\vec{q}_b^\text{bot} = (10 + 30b) \hat{\vec{x}}$ and along the top $\vec{q}_b^\text{top}=\vec{q}_b^\text{bot} + 60\hat{\vec{y}}$ line the channel and break the symmetry. 
The size of the triangles is $\mathcal{R}=20$. 

The orientation of triangles, once again \correctText{}{breaks left-right symmetry and} encourages the MPCD particles to flock from left to right \correctText{}{(the ``easy'' direction).} 
\correctText{and this is indeed seen to be true in the}{Indeed the} majority of realisations \correctText{}{flock in this easy direction} (\fig{fig:VicsekTrianglesChannel}(b)). 
However, in a minority of simulations the flocking goes from right to left --- against the ratchet \correctText{}{in the ``difficult'' direction (\movie{mov:reverse})}. 
The probability of flocking \correctText{in the reverse direction}{in the difficult direction} is approximately $26\%$ \correctText{}{across all densities and activities} (\fig{fig:VicsekTrianglesChannel}(b)). 
As in the previous configuration, the transition to flocking is not as steep as in bulk, suggesting that the obstacles hinder flocking (\fig{fig:VicsekTrianglesChannel}(b)). 

\correctText{}{To better understand the nature of this minority, we run sixty two repeats for $\rho=10$ and $\alpha=0.5$.
We find that $24\%$ flock in the difficult leftward direction, which is consistent with \fig{fig:VicsekTrianglesChannel}(b). 
This suggests that the ratio is relatively independent of the system-wide parameters and most likely set by the geometry of the ratchet channel. 
Furthermore, we observe that once the flock is established, it does not change direction regardless of whether it is flocking in the easy or difficult direction.}
\correctText{Additionally, the flocking speed in the ratchet channel is twice as slow as the wall of funnels (Fig.~17(b)).}{}

These examples illustrate how the LAP-MPCD algorithm can be used to explore the collective motion of active polar fluids in complex geometries, including geometries with an engineered broken symmetry that gets passed on to the spontaneously broken symmetry of the collective flow. 
Future studies could employ LAP-MPCD to study active polar fluids in even more complex geometries~\cite{wamsler2024}, randomised obstacle placement~\cite{keogh2024} or responsive boundaries~\cite{head2024}. 

\section{Conclusion}
We have proposed three new algorithms to simulate active polar fluids based on the Multi-Particle Collision Dynamics (MPCD) framework. 
The first is a hybrid between the MPCD approach and the classic Vicsek model. 
The hybrid Vicsek-MPCD approach reproduces the expectations for the Vicsek model, including exhibiting a flocking transition. 
The critical activity for the flocking transition is found to scale with the density as $\alpha_k \sim \rho^{-1/2}$ in the dilute limit and as $\alpha_k \sim \rho^{-1/3}$ in the dense limit. 
The next two methods we proposed add an active term to the Andersen- and Langevin-thermostatted MPCD algorithms. 
Both of these thermostatted active polar MPCD algorithms exhibit a dynamic phase transition to the flocking state. 
The Langevin active polar MPCD (LAP-MPCD) is explicitly shown to be weakly first order due to the existence of travelling bands near the critical point. 
The group speed of the travelling bands can be measured from kymographs of the density. 
The bands are seen to be stable against perturbations from an external force, such as gravity, until a critical gravity. 
Arrays of asymmetric obstacles slow the transition but do not fully suppress the flocking transition. 
However, the obstacles rectify the flocking direction. 

These results demonstrate that LAP-MPCD \correctText{not only reproduces the key results of the Vicsek model for dry polar active particles but is highly versatile}{captures all the most notable properties of the Vicsek model for dry polar active particles~\cite{Ginelli2015} with the results being consistent with previous studies~\cite{vicsekspp, Chate2008, TonerTu1995}.
Furthermore, the algorithm is highly versatile.}
The MPCD approach has proven to be valuable for simulating passive soft matter systems and is applicable to a wide range of problems, ranging from polymer suspensions~\cite{Zahra}, swimmer dynamics~\cite{zottl2018, mandal2019}, eluting particles\cite{Shendruk}, complex boundary conditions~\cite{keogh2024,wamsler2024}, and much more~\cite{howard2019}. 
Our goal in extending the MPCD approach to simulate active polar fluids is to continue to unlock this same potential for active fluids~\cite{Kozhukhov2022, Benjamin}. 
Continuum models of active polar fluids have been utilised as a model of bacterial suspensions\correctText{ of bacterial suspensions}{}~\cite{Haoran2024, Reinken2020}. 
Active polar MPCD opens a pathway for simulating complex suspensions of bacteria and mesoscale solutes. 
Mesoscale solutes in bacterial suspensions might include small diffusive chemicals like oxygen, large macromolecules such as extracellular DNA, colloids such as microplastics, or droplets formed from biological liquid-liquid phase separation. 
We hope that the methods proposed here will allow future studies to simulate the emergent rheological properties of complex bacteria/solute mixtures including fluctuations without directly solving the hydrodynamical differential equations. 

\section{Acknowledgements}
We thank Timofey Kozhukhov, Zahra Valei, Fran\c{c}ois de Tournemire, Laila Saliekhand, Joseph Knight and 
Ruair\'{i} Phelan for helpful discussions. 
This research has received funding from the European Research Council under the European Union’s Horizon 2020 research and innovation programme (Grant Agreement No. 851196). 
For the purpose of open access, the author has applied a Creative Commons Attribution (CC BY) licence to any Author Accepted Manuscript version arising from this submission.


\dummymov{mov:disordered}
\dummymov{mov:ordered}
\dummymov{mov:bands}
\dummymov{mov:funnels}
\dummymov{mov:ratchetChannel}
\dummymov{mov:reverse}

\bibliography{references}

\end{document}